\documentclass[twocolumn]{aastex631}

\newcommand{\tob}{t_{\rm obs}}
\newcommand{\nuob}{\nu_{\rm obs}}
\newcommand{\nuem}{\nu_{\rm emt}}
\newcommand{\betat}{\Tilde{\beta}}
\newcommand{\gammae}{\gamma^\prime_e}
\newcommand{\revise}{\color{black}}
\newcommand{\reviseJHEAp}{}
\newcommand{\reviseJHEApB}{}

\usepackage[english]{babel}

\usepackage{amsmath}
\usepackage{graphicx}
\usepackage{hyperref}
\usepackage{overpic}
\usepackage[shortlabels]{enumitem}

\begin{document}
\title{The Origin of Cross-Energy-Similar FRED Profiles in Gamma-Ray Bursts Pulses}

\author{S.-X. Yi$^\dagger$}
\affil{State Key Laboratory of Particle Astrophysics, Institute of High Energy Physics, Chinese Academy of Sciences, Beijing 100049, China}
\email{$\dagger$: sxyi@ihep.ac.cn}  
\author{C.-W. Wang}
\affil{State Key Laboratory of Particle Astrophysics, Institute of High Energy Physics, Chinese Academy of Sciences, Beijing 100049, China}
\affil{University of Chinese Academy of Sciences, Chinese Academy of Sciences, Beijing 100049, China}
\author{Shao-Lin Xiong$^\star$}
\affil{State Key Laboratory of Particle Astrophysics, Institute of High Energy Physics, Chinese Academy of Sciences, Beijing 100049, China}
\email{$\star$: xiongsl@ihep.ac.cn}
\author{S.-N. Zhang}
\affil{State Key Laboratory of Particle Astrophysics, Institute of High Energy Physics, Chinese Academy of Sciences, Beijing 100049, China}
\affil{University of Chinese Academy of Sciences, Chinese Academy of Sciences, Beijing 100049, China}
\author{Romain Maccary}
\affil{Department of Physics and Earth Sciences, University of Ferrara, Via Saragat 1, I-44122 Ferrara, Italy}
\affil{INAF - Osservatorio di Astrofisica e Scienza dello Spazio di Bologna, Via Piero Gobetti 101, I-40129 Bologna, Italy}
\author{Rahim Moradi}
\affil{State Key Laboratory of Particle Astrophysics, Institute of High Energy Physics, Chinese Academy of Sciences, Beijing 100049, China}
\author{Shuo Xiao}
\affil{School of Physics and Electronic Science, Guizhou Normal University, Guiyang 550001, People's Republic of China}
\author{Hua Feng}
\affil{State Key Laboratory of Particle Astrophysics, Institute of High Energy Physics, Chinese Academy of Sciences, Beijing 100049, China}
\begin{abstract}
To understand the physical mechanisms underlying the prompt emission of gamma-ray bursts (GRB), single FRED (Fast-Rise-Exponential-Decay) profile GRBs serve as an ideal sample, as they origin from single epoch central engine activity. These GRBs have been found to exhibit a peculiar morphology-including the elegant cross-energy-similarity across energy bands and the recently discovered composite nature—challenging nearly all existing radiation mechanisms, sparking widespread curiosity about their origins. Here we propose a physical model which includes radiation locations sequentially triggered by propagating magnetic perturbations. It naturally explains all observed properties of these GRBs, including the self-similar FRED profile, multi-band aligned subpulses, hard-to-soft spectral evolution, local intensity tracking, and increasing subpulse durations. Furthermore, our results demonstrate that the duration of these GRBs is not reflecting the activity timescale of the central engine, reconciling recent challenges to the traditional merger-short/collapsar-long dichotomy of GRBs.
\end{abstract}

\section{Introduction}
 Although the morphology of gamma-ray bursts (GRBs) exhibit a wide variety, the FRED (Fast-Rise-Exponential-Decay) profile is the fundamental shape of the GRB pulses. This profile is believed to reflect the dissipation and radiation mechanism following a single episode of energy injection from the central engine. In contrast, more complex light curves are thought to result from the superposition of multiple radiation episodes corresponding to the history of central engine activity. In addition, a stochastic-pulse avalanche model, which builds upon the concept of FRED, was developed and accurately reproduces GRB temporal properties, even in the cases of complex multi-pulse light curves \citep{2024A&A...689A.266B,2025A&A...697A..76M}. To understand the physical mechanisms underlying the prompt emission of GRBs, single FRED profile GRBs serve as an ideal sample, as they eliminate the interference from multiple central engine activities. 

FRED-shape GRBs have long attracted the community's attention to their origins from early days due to their ideal morphology of light curves \citep{1992NASCP3137...61K,1996ApJ...459..393N,1996ApJ...473..998F,2005ApJ...627..324N,Liang06,Hakkila08}. Some rather unique properties of such GRB pulses have been discovered since then, e.g., their peaking time delay ($t_{\rm p}$) and pulse width ($t_{\rm w}$) evolve in an energy-dependent manner as power-law, exhibiting softer-later/wider behavior. {\reviseJHEAp This energy-dependent evolution is closely associated with the hard-to-soft spectral evolution \citep{1986ApJ...301..213N,1995ApJ...455L.143S,1997ApJ...485..270S}}. {\reviseJHEApB In the past decade, those spectgral evolution has been widely discussed in the context of the solution of time-evolving electron continuity equation and their synchrotron radiation (e.g., \citealt{2018ApJ...860...72M,2020NatAs...4..174B}); The role of synchrotron self-Compton scattering (SSC) in the electron cooling \citep{2021A&A...656A.134G} and magnetic field fast decay have been discussed \citep{2025A&A...693A.320D}.} Moreover, the power indices of $t_{\rm p}-E$ and $t_{\rm w}-E$ tend to be close or identical \citep{2012ApJ...752..132P}. It means that the profiles in different energy bands are time-stretched copies of each other, {\reviseJHEAp which we refer to as cross-energy similarity \footnote{Note that in \cite{1996ApJ...459..393N}, this phenomenon was refer to as ``a tendency to self-similarity across energy bands'', while in this paper, we use the term ``cross-energy similarity'' to avoid confusion with the term ``self-similar'' used in other contexts.}}. Several types of models have been proposed to explain these properties, such as common curvature effects \citep{2003ApJ...596..389K,2009MNRAS.399.1328G,2012ApJ...752..132P}, emission mechanism \citep{1997ApJ...485..270S,2018ApJ...869..100U,2024ApJ...962...85Y}, and activity history of the central engine \citep{2014styd.confE..90N}. 

Recently, \cite{2025ApJ...985..239Y} demonstrated in the brightest member of this population, \textit{i.e.}  GRB 230307A, that the FRED profile is not elementary but is composed of many short-timescale individual pulses. 
This finding poses a dilemma for GRB models: on one hand, the energy-dependent evolution of the overall profiles indicates that their light curves cannot be attributed to the activity history of the central engine or a series of independent radiation (e.g., a series of internal shocks between pairs of ejecta shells,  \citealt{Kobayashi97,Maxham-Zhang09,2024ApJ...977..155M}); on the other hand, the composite nature of the FRED profile indicates that these GRBs do not arise from a one-go dissipation and radiation as expected from many of the above-mentioned models \citep{1996ApJ...473..998F,2003ApJ...596..389K,2012ApJ...752..132P,2018ApJ...869..100U,2024ApJ...962...85Y}. 

\cite{2025ApJ...985..239Y} suggested that the ICMART (Internal Collision-induced MAgnetic Reconnection and Turbulence, \citealt{zhang2010internal}) framework can reconcile this dilemma, i.e., the light curve profile is composed of emissions from many causally linked local radiation spots. However, \cite{2025ApJ...985..239Y} also pointed out that the conventional ICMART model cannot explain the common cross-energy-similarity in the profiles. To explain the cross-energy-similarity, \cite{2025JHEAp..4700359Y} developed a toy model, in which perturbations propagate outward as concentric circles in the jet front during an ICMART event, triggering local dissipation/radiation points subsequently at different latitudes ({\reviseJHEAp within $1/\Gamma$}). In this toy model, the radiation spectrum is a parameterized phenomenological spectrum. Therefore, we do not consider it a fully self-consistent explanation of the single FRED-shape GRBs. 

In this paper, we build a self-consistent physical model based on the dynamics of the toy model. By solving the electron cooling/radiation equations at local dissipation/radiation points, the energy-time evolution of gamma-ray bursts is derived. Thus, we can reproduce all the energy-time evolution characteristics of single FRED GRBs: including: 1) cross-energy-similar FRED profile; 2) multiband-aligned individual pulses; 3) hard-to-soft spectral evolution; and 4) local intensity tracking. Additionally, we conduct Monte Carlo simulations to establish empirical relationships between physical parameters and observational parameters through population simulations.

The organization of this paper is as follows. In Section 2, we introduce various settings of our model; in Section 3, we demonstrate how our model simulates the various properties of single FRED profile GRBs; Section 4 presents our studies on the empirical relationships of single FRED profiles based on Monte Carlo simulations; finally, we provide our conclusions and discussions. The codes used in this paper are available at \url{https://code.ihep.ac.cn/sxyi/ricarmt}.

\section{The magnetic reconnection model with an expanding perturbation ring}
\begin{figure}
    \centering
    \includegraphics[width=\linewidth]{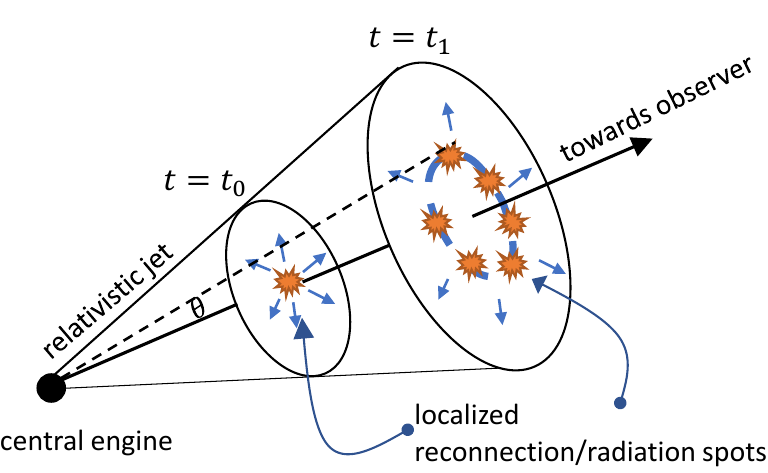}
    \caption{The illustration of the model}
    \label{fig:ill}
\end{figure}

In this model, the central engine impulsively injects energy into a magnetically dominated relativistic jet. At a certain radius, an internal collision within the jet compresses the magnetic field energy into a geometrically thin shell to a critical state, where reconnection can be triggered by magnetic perturbation \citep{zhang2010internal}. Unlike the conventional ICMART model where reconnection seeds are assumed to be triggered simultaneously and spontaneously, here we assume that perturbation and reconnection initially occur in a localized region\footnote{this assumption can be justified in two aspects: 1) the jet shells may have polar anisotropy originating from the central engine, such as a magnetic gradient/density fluctuation at $\theta=0$, favoring local reconnection initiation \citep{2016ApJ...816L..20G,2016MNRAS.459.3635B}; 2) Even macroscopically homogeneous shells have small-scale stochastic magnetic fluctuations, enabling local reconnection initiation \citep{2011ApJ...726...90Z}. One can consider an analogy to phase transition's ``nucleation-growth" process, initiates locally first then propagates globally instead of synchronously across all latitudes.}. This perturbation then propagates {\reviseJHEAp in polar direction} via Alfvén waves across the thin shell, while the shell itself moves forward with the jet at relativistic speed (see illustration in Figure \ref{fig:ill})\footnote{the wave can only propagate within the causally linked region confined by its Mach cone \cite{2012MNRAS.421..570G}, see equation \ref{eq:thetalimit}}. The two-dimensional expansion of the radiation zone, driven by the spreading Alfvén wavefront, accounts for the overall FRED profile of the emission. The expanding wavefront induces further localized magnetic reconnection and radiation in larger ring-shaped areas, corresponding to the individual fast pulses observed in a GRB light curve. {\revise We argue that this successive triggering of the reconnection is physically reasonable, as 1) the scale of the emission region is much larger than the typical magnetic island produced in reconnection; 2) in \cite{uhm2014fast}, a source of persistent injection of accelerated electron is required to reproduce the observed GRB spectrum, which can be naturally provided by the successive triggering of reconnection in our model.}

Under the above settings, we can calculate the observed specific flux of the GRB as follows (see detailed derivation in the {\bf Appendix}):
\begin{widetext}
\begin{equation}
    F_{\nuob}(\tob) = 
    \frac{2\pi\Gamma}{D^2_{\rm{L}}}\int^{t_{\rm max}(t_{\rm obs})}_{t_0}\mathcal{D}^3f(\nuem^\prime,t;t_{\rm inj})R^2\cos\theta\sin\theta\frac{d\theta}{dt_{\rm inj}}dt_{\rm inj},  
    \label{eq:fv_main}
\end{equation}
\end{widetext}

{\reviseJHEAp where $D_{\rm L}$ is the luminosity distance of the GRB; $\Gamma$ is the bulk Lorentz factor of the jet; $\mathcal{D}=[\Gamma(1-\betat\cos\theta)]^{-1}$ is the Doppler factor, with $\betat=\sqrt{1-1/\Gamma^2}$; $R$ is the radius of the emission region; $\theta$ is the polar angle between the line of sight and the local radiation spot; $t_{\rm inj}$ is the time when the electrons begin to inject into the emission site; $t_{\rm max}(t_{\rm obs})$ is the maximum $t_{\rm inj}$ contributing to the observed time $t_{\rm obs}$, which can be derived from the relation between $t_{\rm obs}$ and $t_{\rm inj}$ (see {\reviseJHEAp Appendix}); $f(\nuem^\prime,t;t_{\rm inj})$ is the emissivity function at emission frequency $\nuem^\prime=\nuob(1+z)/\mathcal{D}$ and emission time $t$. From the above equation, we can see overall shape of the light curve profile is determined by the competition between the rising term $R^2\sin\theta$ and the decaying term $\mathcal{D}^3\frac{d\theta}{dt_{\rm inj}}$. The former term increases as the radiation area expands, while the latter term decreases as the Doppler factor decreases with increasing $\theta$ and the $\frac{d\theta}{dt_{\rm inj}}$ decreases as $\theta$ approaches the boundary of the causally linked region (see equation \ref{eq:thetalimit} in {\reviseJHEAp Appendix A}, and also see the figure 2 from the \citealt{2025JHEAp..4700359Y}). The evolution of the emissivity function further modulates the light curve profile in different energy bands.}

{\reviseJHEAp The emissivity function $f(\nuem^\prime,t;t_{\rm inj})$ can be calculated by solving the electron cooling and radiation equations at the local radiation spots. We assume that at $t_{\rm inj}$, electrons are accelerated into a power-law distribution with index $p$ between $\gamma_m$ and $\gamma_M$. The magnetic field strength in the comoving frame of the jet decreases with radius as $B^\prime=B_0^\prime (R/R_0)^{-1}$, where $B_0^\prime$ is the magnetic field strength at radius $R_0$. The electrons cool via synchrotron radiation and adiabatic expansion. The detailed calculation of $f(\nuem^\prime,t;t_{\rm inj})$ is presented in the {\bf Appendix A}.}

\section{Reproducing the spectrum-time features of individual bursts}\label{sec:reproduce}
With $F_{\nuob}(\tob)$ known, the light curve of a GRB in different energy bands can be obtained by integrating the specific flux over the corresponding energy band. The spectrum of a GRB at a certain time interval can be obtained by averaging $F_{\nuob}(\tob)$ over the time interval. In the above derivation, we assume the emissivity of the GRB is smooth. {\reviseJHEApB To simulate this discrete nature, we replace the continuous integral over $t_{\rm inj}(\theta)$ in Equation (\ref{eq:fv}) with a summation over a finite number ($N_{\rm ring}$) of radiation polar angles (rings). The injection of electrons within each ring is not instantaneous; rather, it has a finite duration $\Delta t_{\rm inj}$. This duration is determined by the crossing time of the propagating wave between successive rings: $\Delta t_{{\rm inj},i} = t_{{\rm inj},i+1}(\theta_{i+1}) - t_{{\rm inj},i}(\theta_i)$. Within this duration, the injection rate is assumed to be uniform. The latitude of the rings, $\cos\theta_i$, is uniformly randomly distributed in the range of $[\cos\theta_{\rm max},1]$.}

\begin{figure}
    \centering
    \includegraphics[width=0.5 \textwidth]{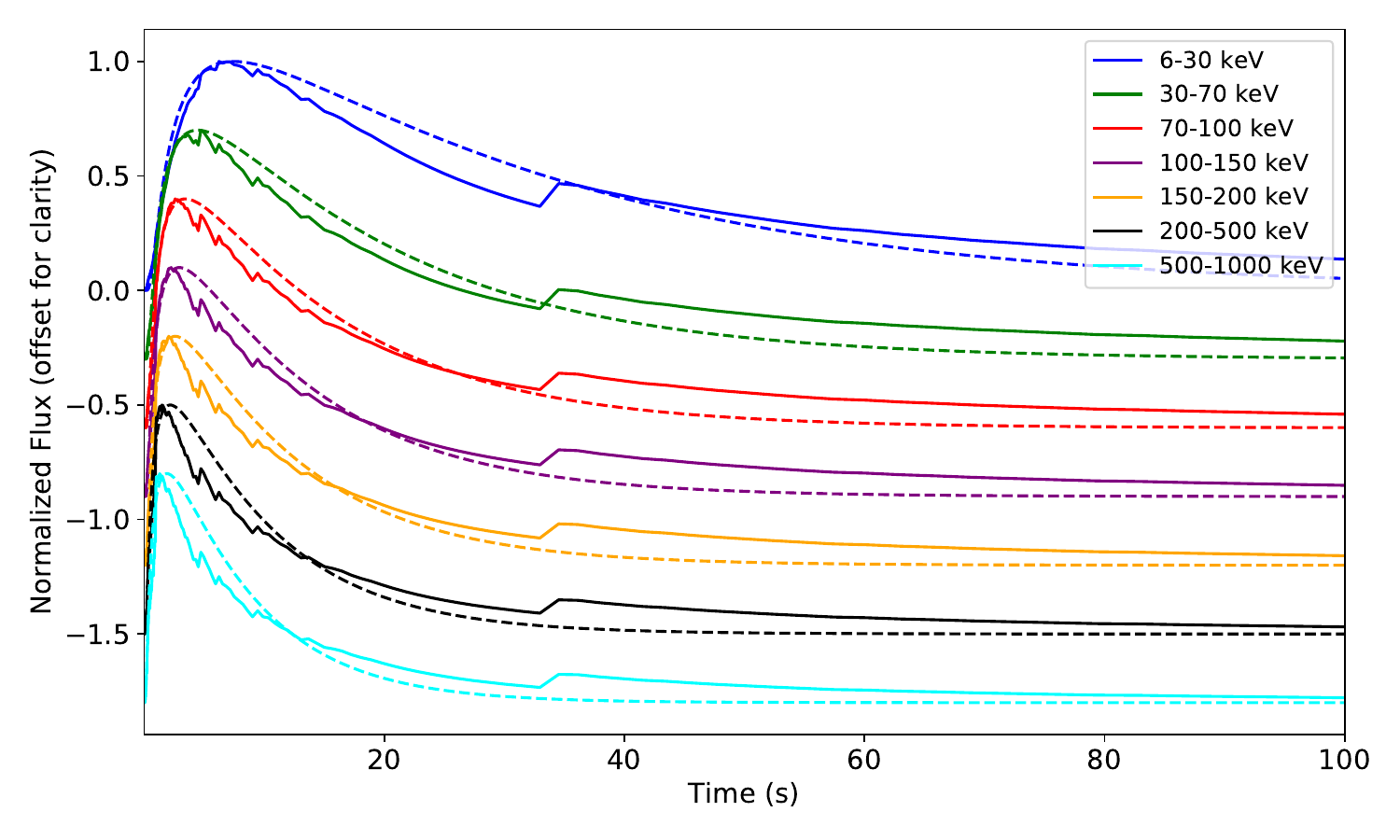}
    \caption{The simulated light curve of a GRB in 7 different energy bands, i.e., 6-30keV, 30-70keV, 70-100keV, 100-150keV, 150-200keV, 200-500keV and 500-1000keV.}
    \label{fig:lc}
\end{figure}
\begin{figure}
    \centering
    \includegraphics[width=0.5\textwidth]{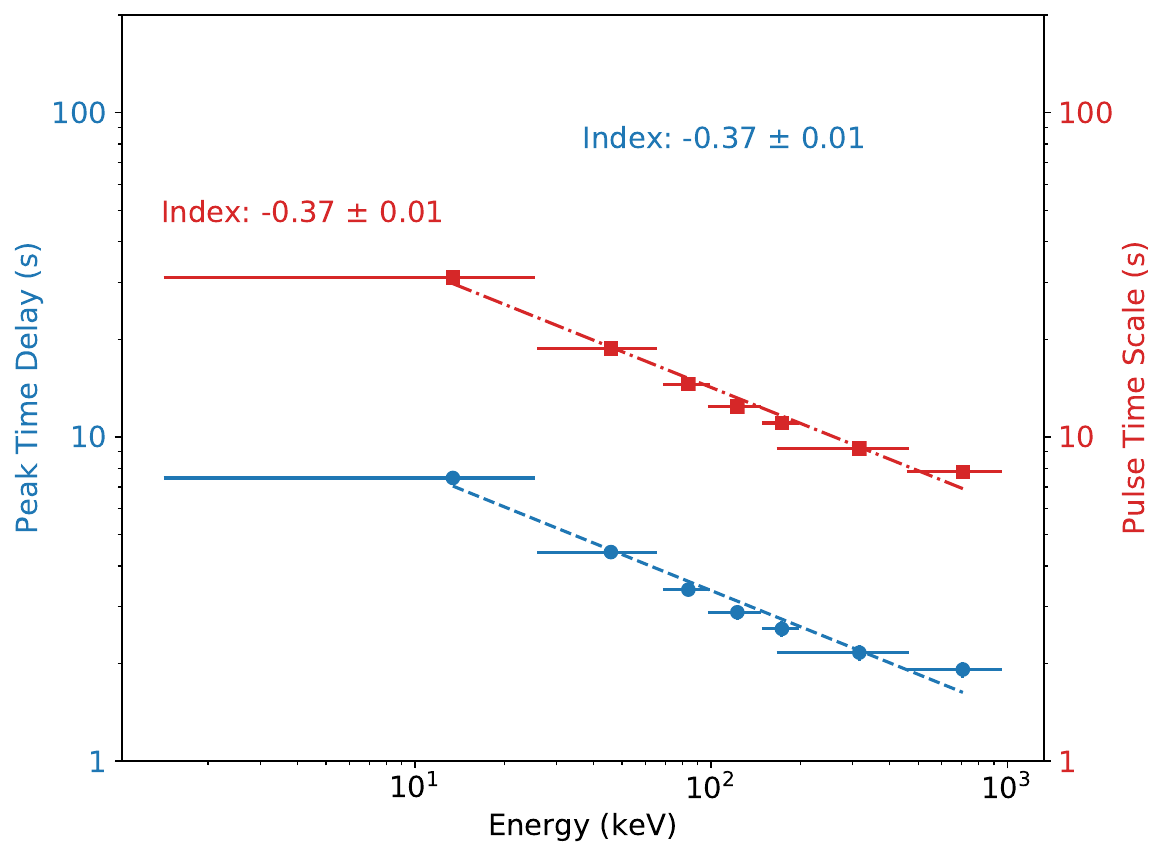}
    \caption{The peak time and width of the simulated GRB light curves as a function of energy. The best fit power law are plotted as dashed lines.}
    \label{fig:tw}
\end{figure}

As an example, we take the following parameters to simulate a GRB with our model: $\Gamma=180$, $r_0=2\times10^{15}$cm, $B_0^\prime=1$G, $p=2.8$, $\gamma_m=2\times10^5$, $\gamma_M=10^7$, $z=0.06$, $\Delta r^\prime=10^{10}$cm and $N_{\rm ring}=200$. The choice of model parameters in our simulation is motivated by both observational constraints and theoretical considerations. The Lorentz factor $\Gamma$, initial radius $r_0$, and magnetic field $B_0'$ are selected to be representative of typical values. The electron distribution parameters ($\gamma_m$, $\gamma_M$, $p$) are chosen to reflect the range commonly required to reproduce observed GRB spectra \citep{uhm2014fast}. The number of radiation rings $N_{\rm ring}$ is set to balance the observed diversity in light curve smoothness and variability. Our results are robust to moderate changes within a reasonable ranges, and the qualitative features of the model persist. In Figure \ref{fig:lc}, we show the simulated light curve of the GRB in 7 different energy bands, i.e., 6-30 keV, 30-70 keV, 70-100 keV, 100-150 keV, 150-200 keV, 200-500 keV and 500-1000 keV. The light curves exhibit a typical FRED profile, with the peak time and width evolving with energy. We fit the light curves with a FRED function defined in \cite{2005ApJ...627..324N}: 
\begin{equation}
    F(t)\propto\exp\left(-(\frac{t-t_0}{t_d}+\frac{t_r}{t-t_0})\right).
    \label{equ:norris05}
\end{equation}
In this formulation, we define the peaking time delay as $t_p=\sqrt{t_rt_d}$, and the pulse time scale (a simple proxy of the pulse width) as $t_w=t_r+t_d$. The fitted $t_p$ and $t_w$ as a function of energy are shown in Figure \ref{fig:tw}. The value of the energy is the geometrical mean of each energy band, while the error bar of energy is the width of the energy band. The best fit FRED profiles are plotted along with the light curves in Figure \ref{fig:lc}. 

It can be seen that the energy dependent of $t_p$ and $t_w$ can be fitted with a power law, with the fitted power indices being both $-0.37$. 

\begin{figure}
    \centering
    \includegraphics[width=0.5 \textwidth]{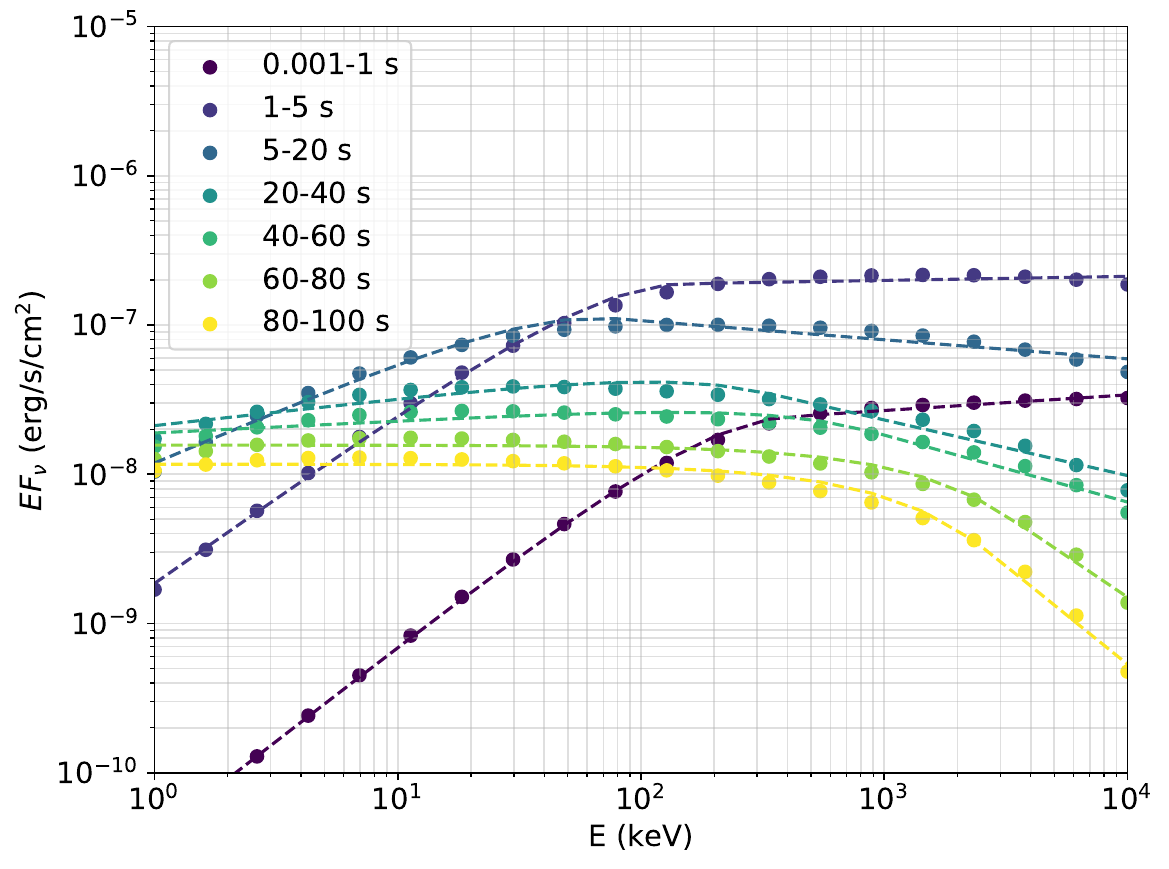}
    \caption{\revise The simulated GRB spectra at 7 different time windows (in dots), along with the best fit Band formulation in dashed lines.}
    \label{fig:spec}
\end{figure}

\begin{figure}
    \centering
    \includegraphics[width=0.5 \textwidth]{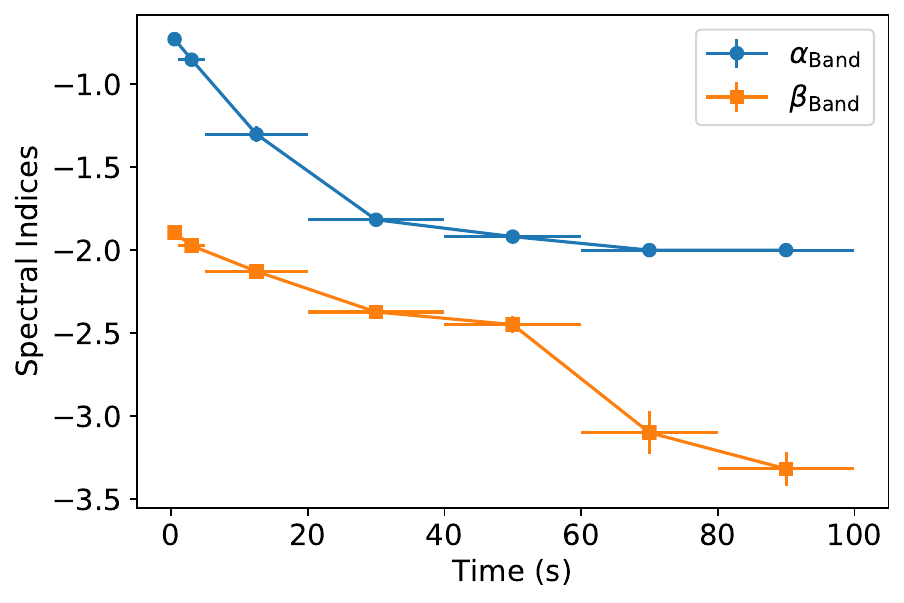}
    \caption{\revise The best-fit low-energy ($\alpha_{\rm Band}$) and high-energy ($\beta_{\rm Band}$) Band spectral indices of the simulated GRB spectra as a function of time.}
    \label{fig:Band-indice}
\end{figure}

The corresponding spectrum of the GRB at 7 different time windows are shown in figure \ref{fig:spec}. The spectra exhibit a typical Band function shape, with the peak energy $E_p$ evolving with time. {\revise To better illustrate this, we fit the spectra with the Band function \citep{1993ApJ...413..281B}, and plot the best fit Band functions along with the simulated spectra in figure \ref{fig:spec}. In figure \ref{fig:Band-indice}, we plot the best-fit low-energy ($\alpha_{\rm Band}$) and high-energy ($\beta_{\rm Band}$) spectral indices as a function of time, in which it can be seen that both indices evolve from hard to soft with time, consistent with the typical spectral evolution pattern of GRBs.}

\begin{figure}
    \centering
    \includegraphics[width=0.5\textwidth]{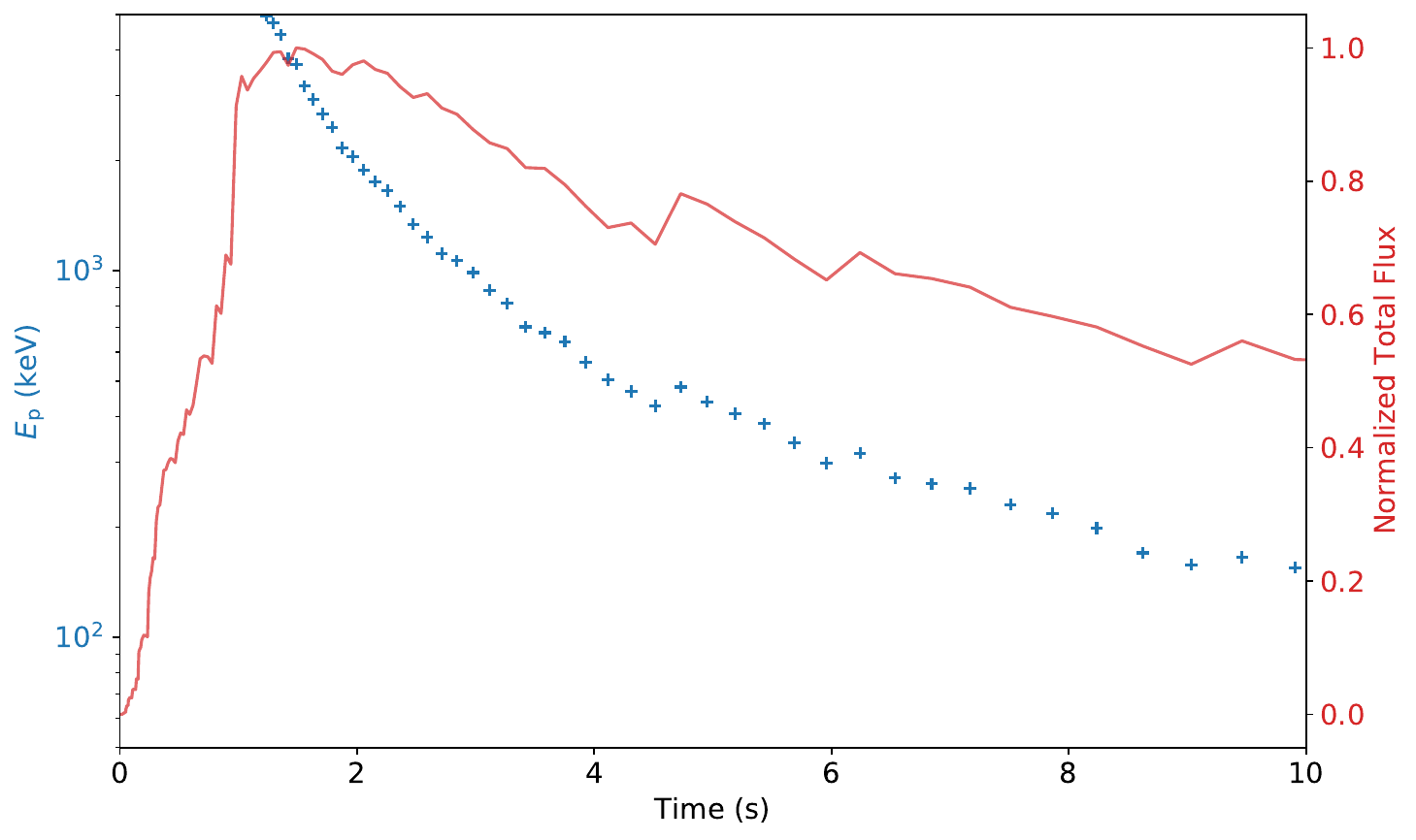}
    \caption{The $E_p$ as function of time, along with the energy integrated light curve of the simulated GRB.}
    \label{fig:it}
\end{figure}

It is worth mentioning that not only the overall cross-energy-similar FRED profile can be reproduced, but also it is clearly shown that these FRED profile are composed of many individual pulses, which are aligned in different energy bands, as in the observation of GRB 230307A \citep{2025ApJ...985..239Y}. These individual pulses correspond to the localized radiation spots in the random discrete rings. The duration of these individual pulses also shows systematic trend of increasing with time, which is consistent with the recent findings in GRB 230307A \citep{2025arXiv250905628M}. In our model, this trend is a natural outcome of the decreasing of the Doppler factor of individual radiation rings with time, as the radiation area propagates outward to higher latitudes. Furthermore, we also simulate the instantaneous spectrum $\nu f_{\rm\nu}$ of the simulated GRB and find the evolution of $E_p$. We plot $E_p-t$ along with the energy integrated light curve in figure \ref{fig:it}. It can be seen that, in addition to the general decrease in long time scale, $E_p$ tracks the intensity of the GRB in short time scale, which is common in GRBs and also consistent with the observation of GRB 230307A \citep{sun2023magnetar}. This local intensity tracking can be explained in our framework as the cooling of the electrons in the local radiation spots. 

\section{The empirical relationships between physical parameters and observational parameters of single FRED GRBs}
\begin{figure}
    \centering
    \includegraphics[width=0.5\textwidth]{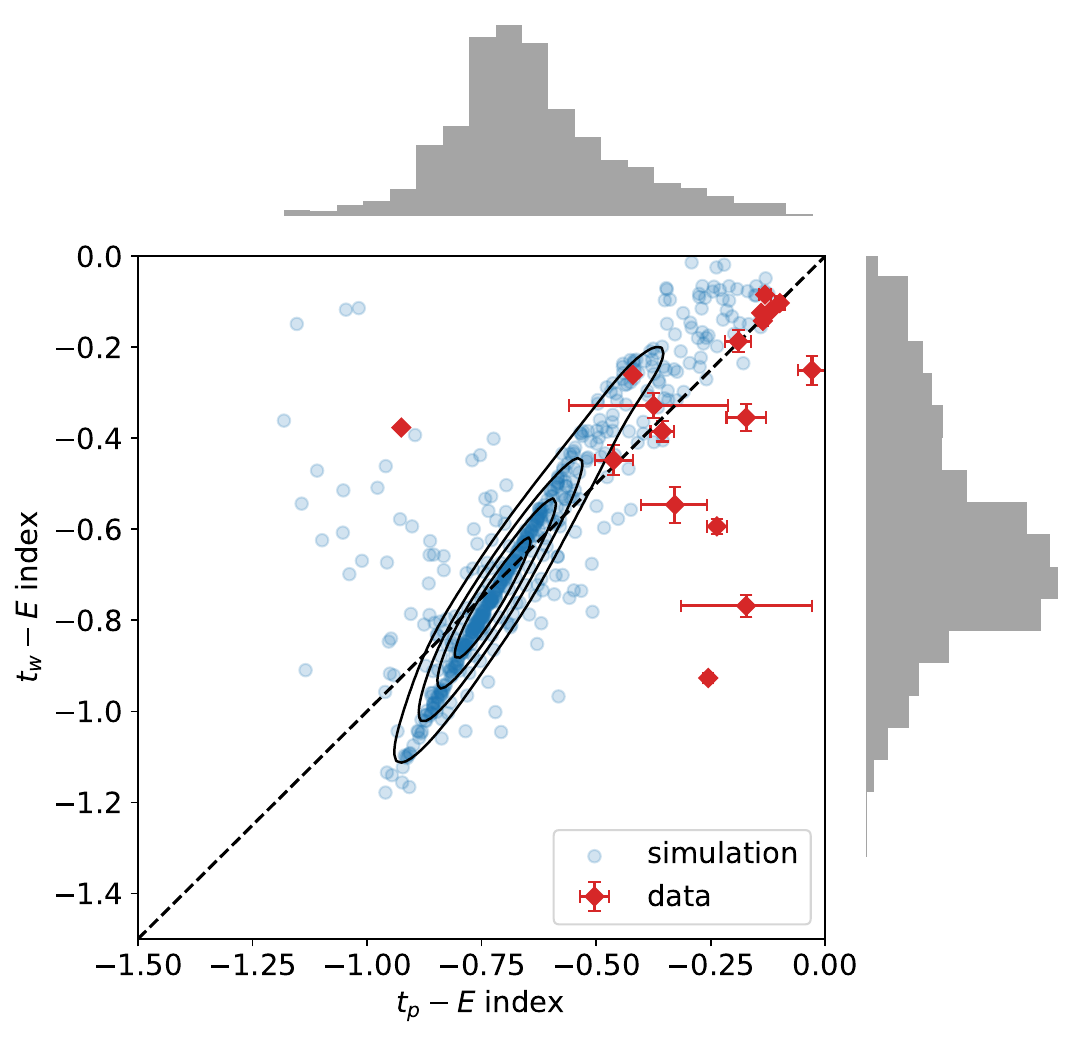}
    \caption{The distribution of the $t_p-E$ and $t_w-E$ power indices of the simulated population of single FRED GRBs. The dashed line is the diagonal line where the two indices are identical. The red points are the observed parameters from a new catalogue of single FRED GRBs (See details in the {\bf Appendix: a new catalogue of general single FRED GRBs}).}
    \label{fig:pop-tw}
\end{figure}

\begin{figure}
    \centering
    \includegraphics[width=0.5\textwidth]{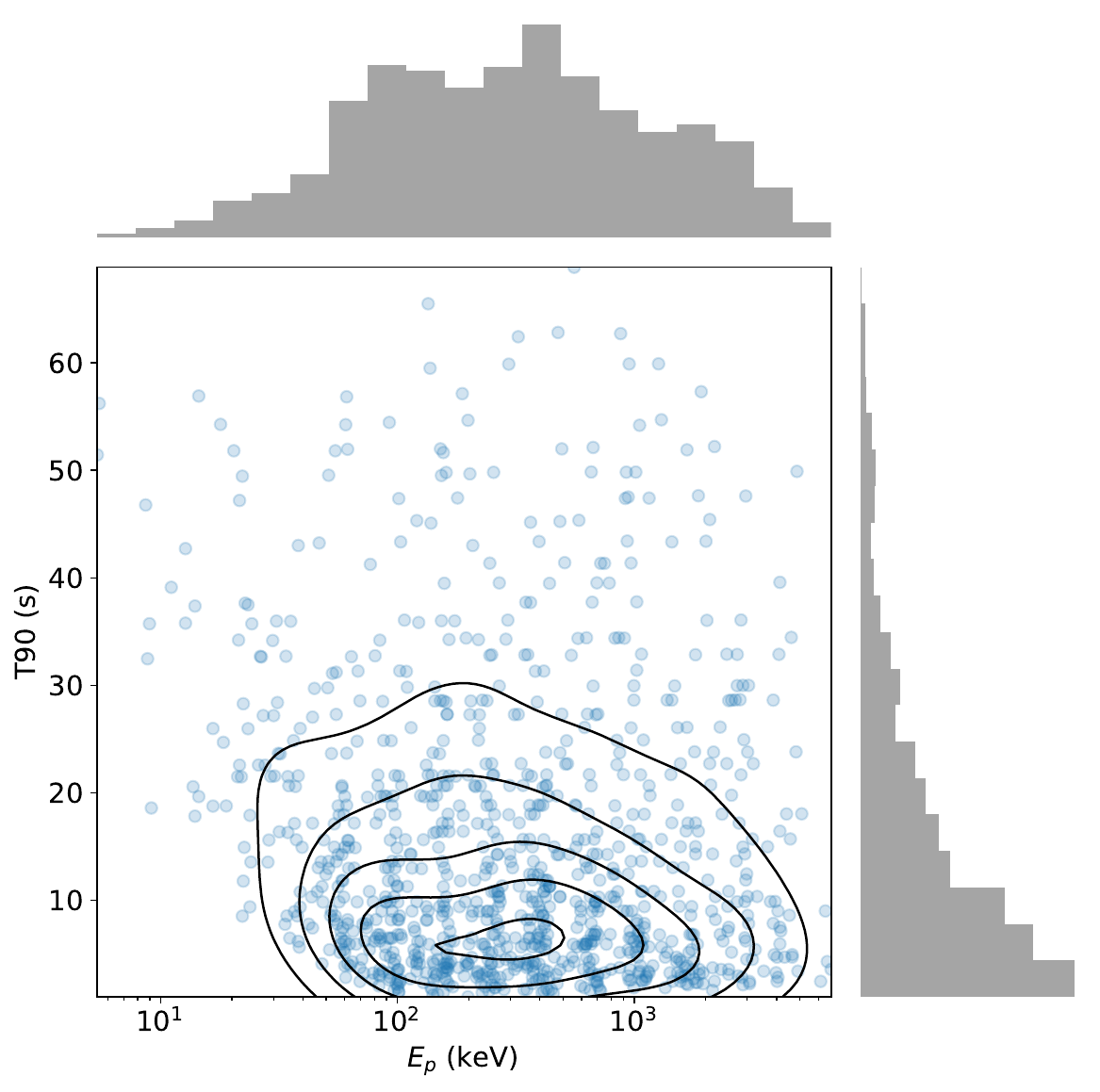}
    \caption{The distribution of T$_{90}$ and $E_p$ of the simulated population of single FRED GRBs.}
    \label{fig:pop-te}
\end{figure}

Given a set of model parameters, we can accurately reproduce the spectrum-time features of a single FRED GRB as demonstrated in the previous section. With a distribution of model parameters, we can further generate a population of such GRBs. The distribution of the model parameters are set as follows: $\Gamma$ is uniformly distributed from 100 to 500; $r_0$ is uniformly distributed in log-space from $5\times10^{14}$ cm to $5\times10^{15}$ cm; $B_0^\prime$ is uniformly distributed from 3 to 30 G. {\reviseJHEApB The range of $\Gamma$, $B_0$ and $r_0$ are consistent with the tentative constraints in earlier studies \citep{2025ApJ...986..106G}, although it is still challenging to constrain those physical parameters directly from observation;}  $\gamma_m$ is log-uniformly distributed from $10^4$ to $10^5$, $\gamma_M$ is log-uniformly distributed from $10^7$ to $10^8$. {\reviseJHEApB These parameters are consistent with the range choosen in earlier simulation work (e.g., \citealt{uhm2014fast})}. From the simulated population of size 1000, we calculate the T$_{90}$ of the energy-integrated light curve, the peak energy $E_p$ of the time averaged spectrum, and $t_p-E$ and $t_w-E$ power indices. In figure \ref{fig:pop-tw}, we show the distribution of the $t_p-E$ and $t_w-E$ power indices. It can be seen that the indices are clustered around the diagonal line, where the both indices are identical, corresponding to perfect cross-energy-similar profiles. Therefore, a cross-energy-similar FRED of such GRBs is a natural outcome of our mechanism. The T$_{90}$ and $E_p$ of the simulated population are shown in figure \ref{fig:pop-te}. It can be seen from the distribution of T$_{90}$ that, a lot of the single FRED GRBs are short duration GRBs, consistent with the impulsive central engine activity expected for Type-I GRBs. Still, there are some long duration GRBs with T$_{90}$ of several tens of seconds, which explains the existence of long duration Type-I bursts such as GRB 211211A and 230307A \citep{2022Natur.612..223R,2024Natur.626..737L,typeIL_wang_2025,typeIL_tan_2025}. In figures \ref{fig:pop-tw}, we also superpose the observed parameters from a set of recently discovered single FRED GRBs (See details in the {\bf Appendix: a new catalogue of general single FRED GRBs}). As can be seen, the data points are also clustered around the diagonal line, indicating cross-energy-similar profiles, which is consistent with our simulation results. However, there seems to be an excess of outliers with $t_p-E$ index larger than $t_w-E$ index. This may suggest a different distribution of the physical parameters than the one we assumed in our simulation. More data and more detailed analysis are needed to clarify this issue in the future.

We further fit an empirical relation between the physical parameters and the observational parameters of the simulated population. The empirical relation is defined as: 
\begin{equation}
    \Vec{X}_{\rm obs}= \Vec{Y}_{\rm phys}\mathbf{A}+\Vec{B},
\end{equation}
where 
\begin{equation}
    \Vec{X}_{\rm{obs}}=[\log_{10}E_p,\log_{10}\text{T}_{90},  \text{index}_{t_p}, \text{index}_{t_w}],
\end{equation}
\begin{equation}
    \Vec{Y}_{\rm{phys}}=[\log_{10}\Gamma, \log_{10}r_0, \log_{10}B_0^\prime, \log_{10}\gamma_m, \log_{10}\gamma_M],
\end{equation}
$\mathbf{A}$ is a $4\times5$ matrix defining the linear coefficients, and $\Vec{B}$ is a 4-dimensional vector of the intercepts. The best fit coefficients are:

\begin{equation}
\mathbf{A} =
\begin{pmatrix}
1.3 & -0.2 & 0.3 & 1.7 & -0.007 \\
-1.5 & 0.9 & 0.1 & 0.3 & 0.006 \\
-0.4 & 0.1 & -0.3 & -0.3 & -0.003 \\
-0.6 & 0.2 & -0.3 & -0.3 & -0.004 \\
\end{pmatrix}
\end{equation}

\begin{equation}
\Vec{B} =
\begin{pmatrix}
-5.1 \\
-9.8 \\
0.5 \\
-0.4 \\
\end{pmatrix}
\end{equation}

\begin{figure*}
    \centering
    \includegraphics[width=\textwidth]{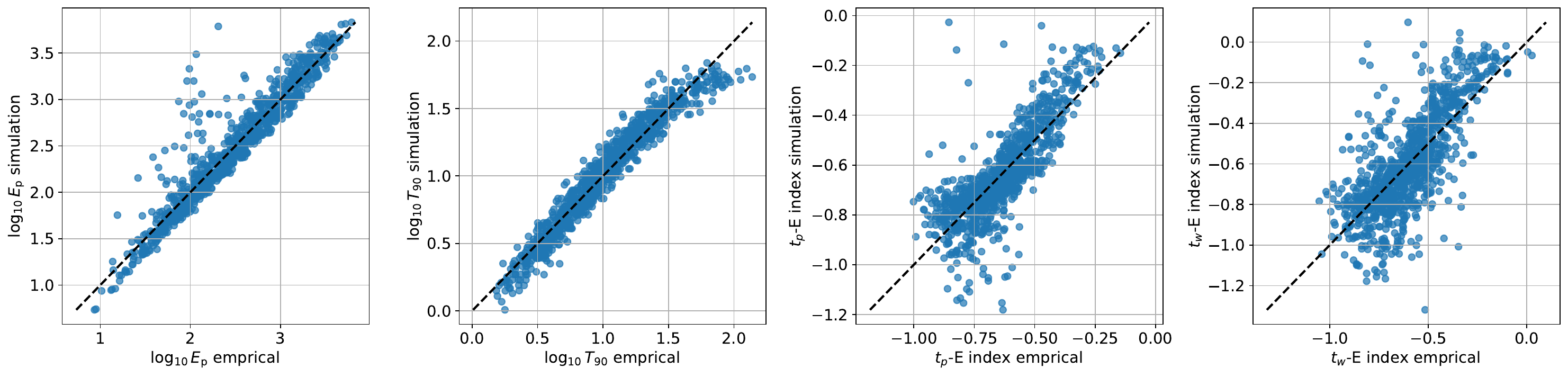}
    \caption{The comparison between the observables calculated with the empirical relation and the ones calculated with the simulations. The dashed lines are the diagonal lines where the two values are identical.}
    \label{fig:empirical}
\end{figure*}

In order the estimate the goodness of the empirical relation, we plot the observables calculated with the empirical relation against the ones calculated with the simulations in figure \ref{fig:empirical}.

The empirical relation can be used to estimate the observables of a single FRED from our scenario with a set of physical parameters with much less computational cost. It can also be used to estimate the physical parameters of a single FRED GRB with its observables. 

\section{conclusions and discussions}
In this paper, we have established a self-consistent physical model that attributes the peculiar cross-energy-similar FRED profile in GRBs, to the propagation of internal collision-caused perturbation in a magnetically dominated relativistic jet front. The initial perturbation are caused by an impulsive energy injection from the central engine and is localized. The perturbation then propagates outwards in a geometrically thin shell in the jet front, in the form of Alfvén waves. The perturbation propagates in concentric circles, and randomly triggers local magnetic reconnection and radiation at different latitudes, which results in the overall FRED profile of the GRB. The local radiation spots are also responsible for the individual pulses in the light curve, which are aligned in different energy bands. 

In each localized radiation spot, energetic electrons are accelerated by the magnetic reconnection and then endures cooling from synchrotron radiation and adiabatic expansion as the ejecta moving forward. We derive the energy-time evolution of the GRB by solving the electron cooling and radiation equations at each local radiation spot. 

All the features of single FRED GRBs can be reproduced with our model, including: 1. cross-energy-similar FRED profile, which is due to the concentric circle propagation of the perturbation; 2. multi-band aligned individual pulses, which is due to the random distribution of the local radiation spots; 3. overall hard-to-soft spectral evolution, which is due to the decreasing Doppler factor of the radiation spots as they propagate outward; 4. local intensity tracking, which is due to the cooling of the electrons in the local radiation spots. 5. A systematic trend of increasing duration of the individual pulses, which is due to the decreasing Doppler factor of the radiation spots as they propagate outward. 

Furthermore, we conduct Monte Carlo simulations to study the population properties of such GRBs, given a distribution of physical parameters. We find that the $t_p-E$ and $t_w-E$ power indices are clustered around the diagonal line, which indicates that the cross-energy-similar FRED profile is a natural outcome of our model. The T$_{90}$ and $E_p$ of the simulated population are consistent with the observed single FRED GRBs, including both short and long duration bursts. We also establish an empirical relation between the physical parameters and the observables of single FRED GRBs, which can be used to estimate the observables from a set of physical parameters with much less computational cost, or vice versa.

In the observation, some single FRED GRBs exhibit smooth light curves, while others show high short time variability. In our model, the smoothness of the light curve is controlled by the number of local radiation spots. More radiation spots result in a smoother light curve, while fewer radiation spots lead to a more variable light curve. The variability can be further increased if there are minijets within each local radiation spot, as shown in the simulation studies \citep{2014ApJ...782...92Z,2022ApJ...927..173S}. 

In our model, the reproduction of a canonical FRED shape depends on two assumptions: 1. there is only one initial perturbation; 2. the initial location of the perturbation is close to the line of sight. If the initial perturbation is away from the line of sight, the profile will be distorted from the canonical FRED shape; if there are multiple initial perturbation on the other hands, the profile will be more complex, being the superposition of multiple distorted FRED profiles.

{\reviseJHEAp In the derivation of the emissivity, the cooling of the electron is only considered to be from synchrotron radiation and adiabatic expansion and we ignored the enhenced cooling from the SSC, which could be significant in some cases \cite{2001ApJ...548..787S,2009ApJ...703..675N}. We leave the detailed study of the SSC effect in our model to future work.} The radiation mechanism in this model is assumed to be pure synchrotron radiation. In reality, if the photons are partially thermalized when the radiation location is compact, the spectrum will be softer at the beginning of the burst, as observed in GRB 230307A \citep{sun2023magnetar}.  

From our study, it is clearly shown that the cross-energy-similar FRED profile of GRBs is reflecting the dissipation within the jet, rather than the activity history of the central engine. Therefore, the recent changing to the traditional merger-short/collapsar-long dichotomy of GRBs can be reconciled \citep{2025JHEAp..45..325Z}. It is also elaborated why we believed that the composed cross-energy-similar FRED profile is a strong evidence of magnetically dominated jets. 

There has been efforts to infer the central engine activity from the light curve of GRBs. However, as we pointed out here, the light curve of GRBs is not a direct reflection of the central engine activity, but rather a convolution of the response to the impulsive central engine activity and the history of the central engine activity. As a result, knowing the properties of such response is criucial for understanding the central engine activity. From a series of our studies \citep{2025ApJ...985..239Y,2025JHEAp..4700359Y}, we have shown that the cross-energy-similar FRED profile is such a response. Therefore, the long time activity history of the central engine can be accurately studied.  

{\reviseJHEAp Finally, we would like to extend the dicussion between the soften-wider/later phenomena and the spectral-lag. Both descriptions are phenomenological ways to characterize the energy-time evolution of GRBs: the former discription focuses on the temporal profile of the light curve in different energy bands, while the latter two focus on the spectral evolution characteristics. The equivalence between these two descriptions is straightforward. However, a trivial spectral-lag cannot lead to the observed trend of cross-energy similarity. Although our model motivated by the light curve across energy, in our model, the evolution of the energy spectrum and the lightcurve evolution caused by propagating dynamics are equally fundamental in shaping the cross-energy pulse profile.
}

\begin{acknowledgments}
We thank the insightful discussion with Prof. Bing Zhang and Dr. Maria Ravasio. This work is supported by the Chinese Academy of Sciences Institute of High Energy Physics Innovation Project (E3545KU2), the National Natural Science Foundation of China (Grant No. 12333007) and China's  Space Origins Exploration Program. 
\end{acknowledgments}
\appendix
\section{A detailed derivation of the observed specific flux $F_{\nuob}(\tob)$ from an expanding ring of magnetic reconnection and radiation}
Under the our model scenario, the observed specific flux of a GRB is (see derivation in the toy model paper, \citealt{2025JHEAp..4700359Y}):
\begin{equation}
    F_{\nuob}(\tob)=\frac{4\pi\Gamma}{D^2_{\rm{L}}}\int\int\mathcal{D}(\theta)^3 j^\prime_{\nuem^\prime}(r,t,\theta)\sin\theta d\theta r^2dr,
\end{equation}
where $\Gamma$ is the Lorentz factor of the jet, $\mathcal{D}$ is the Doppler factor of the emitters, $j^\prime_{\nu^\prime}$ is the emissivity in the co-moving frame, $t$ is the time of emission in the central engine-rest frame, and $t_{\rm obs}$ is the time in the observer frame. $t$ is a function of $\tob$ and $\theta$ as:
\begin{equation}
t=\frac{\tob}{(1+z)(1-\beta\cos\theta)}+t_0,
\end{equation}
where $t_0=r_0/(\beta c)$ and $r_0$ is the radius of the initial reconnection, $\beta=v/c$ is the velocity of the jet with respect to the speed of light. 

According to the dynamics of our model, the emissivity has the form:
\begin{equation}
    j^\prime_{\nuem^\prime}(r,t,\theta)=\delta(r-R(t))f(\nu,t,t_{\rm inj}(\theta)),
\end{equation}
where $f(\nu^\prime,t,t_{\rm inj}(\theta))$ denotes the spectrum of emitted photons at centre engine rest frame time $t$ due to electrons injected at $t=t_{\rm inj}$ at a certain ring $\theta$. The relation between $t_{\rm inj}$ and $\theta$ is (see the toy model paper, \citealt{2025JHEAp..4700359Y}):
\begin{equation}
    t_{\rm inj}=\frac{t_0}{1-\theta\beta\Gamma/\Tilde{\beta}}.
\end{equation}
Since $f(\nuem^\prime,t;t_{\rm inj})$ is nonzero only at $t\ge t_{\rm inj}$, and there is one-to-one function between $\theta$ and $t_{\rm inj}$, the integral over $\theta$ can be rewritten as integral over $t_{\rm inj}$ to a maximum value $t_{\rm max}$ corresponding to $\tob$.
the above equation become:
\begin{equation}
    F_{\nuob}(\tob) = 
    \frac{2\pi\Gamma}{D^2_{\rm{L}}}\int^{t_{\rm max}(t_{\rm obs})}_{t_0}\mathcal{D}^3f(\nuem^\prime,t;t_{\rm inj})R^2\cos\theta\sin\theta\frac{d\theta}{dt_{\rm inj}}dt_{\rm inj}.   
    \label{eq:fv}
\end{equation}

In the above equation, the integration upper limit $t_{\rm max}$ is solely determined by $\tob$. 
The relation between $t_{\rm max}$ and $\tob$ is found with the following two equations:
\begin{eqnarray}
\tob&=&(1+z)(1-\beta\cos{\theta})(t_{\rm max}-t_0) \\
\theta&=&\frac{1-t_0/t_{\rm max}}{\beta\Gamma/\Tilde{\beta}}\label{eq:thetalimit}.
\end{eqnarray} 
The Doppler factor $\mathcal{D}$ is a function of $\theta$, and the latter is a function of $t_{\rm inj}$:
\begin{equation}
    \theta=\frac{1-t_0/t_{\rm inj}}{\beta\Gamma/\Tilde{\beta}}.
\end{equation}
$\nu^\prime$ is determined by $\nu_{\rm obs}$ and $\mathcal{D}$, $\nu^\prime=\nuob/((1+z)\mathcal{D})$, therefore also a function of $t_{\rm inj}$ that takes place in the integral. 
$R$ is function of $t$:
\begin{equation}
    R=\beta ct.
\end{equation}
However $t$ is also determined by $\tob$ and $\theta$ as:
\begin{equation}
    t=t_0+\frac{\tob}{(1+z)(1-\beta\cos\theta)}.
\end{equation}
Since $\theta$ has a one-to-one mapping to $t_{\rm inj}$, $t$ is a function of $\tob$ and $t_{\rm inj}$: $t(\tob,t_{\rm inj})$. 

The key is to find the formula of $f(\nuem^\prime,t;t_{\rm inj})$. In order to do that, we need to get the spectrum of the electrons $dn_e(\gamma^\prime_e, t; t_{\rm inj})/d\gamma^\prime_e$, where $\gamma^\prime_e$ is the Lorentz factor of the electron in the co-moving frame. 

We assume that the $\gammae$ follows a power law at injection instant:
\begin{equation}
    dn_e(\gammae, t_{\rm inj}, t_{\rm inj})/d\gammae = \dot{n}_{\rm e}(t_{\rm inj})\kappa{\gammae}^{-p}\delta t_{\rm inj},
    \label{eq:t0inject}
\end{equation}
where $\dot{n}_{\rm e}(t_{\rm inj})$ is the electron injection rate density. In the above equation, $\kappa$ is the normalization factor of the probability density so that:
\begin{equation}
    \int_{{\gamma}_m}^{\infty} \kappa{\gammae}^{-p}=1,
\end{equation}
where $\gamma_m$ is the minimum electron Lorentz factor at injection. By normalizing the power-law distribution, $\kappa=\frac{p-1}{{\gamma}_m^{1-p}}$. The kinetic energy density of the injected electrons are ($p>2$):
\begin{equation}
    \epsilon =\kappa\dot{n}_e\delta t_{\rm inj}m_ec^2\int^{\infty}_{{\gamma}_m}{\gammae}^{1-p}d\gammae=\frac{p-1}{p-2}\gamma_m\dot{n}_e\delta t_{\rm inj}m_ec^2.
\end{equation}
In this scenario, the electrons are accelerated by the reconnection of the local toroidal magnetic field. 
Therefore:
\begin{equation}
    \epsilon=\frac{{B^\prime}^2}{4\pi},
\end{equation}
where $B^\prime$ is the magnetic field strength in the co-moving frame. According to the conservation of the toroidal magnetic field energy, $B$ scaled with $1/r$ as:
\begin{equation}
    B^\prime=B_0^\prime\frac{r_0}{r}. 
\end{equation}

As a result, $\dot{n}_e$ can be found out to be:
\begin{equation}
   \dot{n}_e = \frac{{B^\prime}^2}{4\pi m_ec^2\delta t_{\rm inj}}\frac{p-2}{p-1}\frac{1}{\gamma_m}.
\end{equation}
For each electron with Lorentz factor $\gammae$, its energy evolution follows \citep{uhm2014fast}:
\begin{equation}
    \frac{d}{dt^{\prime}}y=\frac{\sigma_T}{6\pi m_ec}{B^\prime}^2-\frac{1}{3}y\frac{d\ln n_e}{dt^{\prime}},
    \label{eq:cooling}
\end{equation}
where $y\equiv1/\gammae$. The first term on the right-hand side of the above equation corresponds to the synchrotron cooling, while the second term corresponds to the adiabatic cooling. Since $n_e\propto r^{-2}$, the above adiabatic term becomes:
\begin{equation}
    \frac{dy}{dt^\prime}|_{\rm adiab} = \frac{2\beta c\Gamma}{3}\frac{y}{r}.
    \label{eq:evolution}
\end{equation}

Following \cite{uhm2014fast}, we find $dn_e(\gamma^\prime_e, t; t_{\rm inj})/d\gamma^\prime_e$ by the following method:
\begin{enumerate}
    \item In an array of $\gammae$ bins from $\gamma^\prime_m$ to $\gamma^\prime_M$, we assign the number of electrons according to equation (\ref{eq:t0inject});
    \item For each bin at $\gammae$, we evolve the energy of electron according to equation (\ref{eq:cooling}) to time $t$;
    \item At time $t$, we redistribute the new $\gammae(t)$ into the bins, and assign weight according to the number of electrons in its initial bin.
\end{enumerate}

{\reviseJHEAp In figure \ref{fig:dndg}, we plot the evolution of electron spectrum at different time since injection from different injection latitudes, with parameters listed in the section \ref{sec:reproduce}.} 

{\begin{figure}
    \centering
    \includegraphics[width=\textwidth]{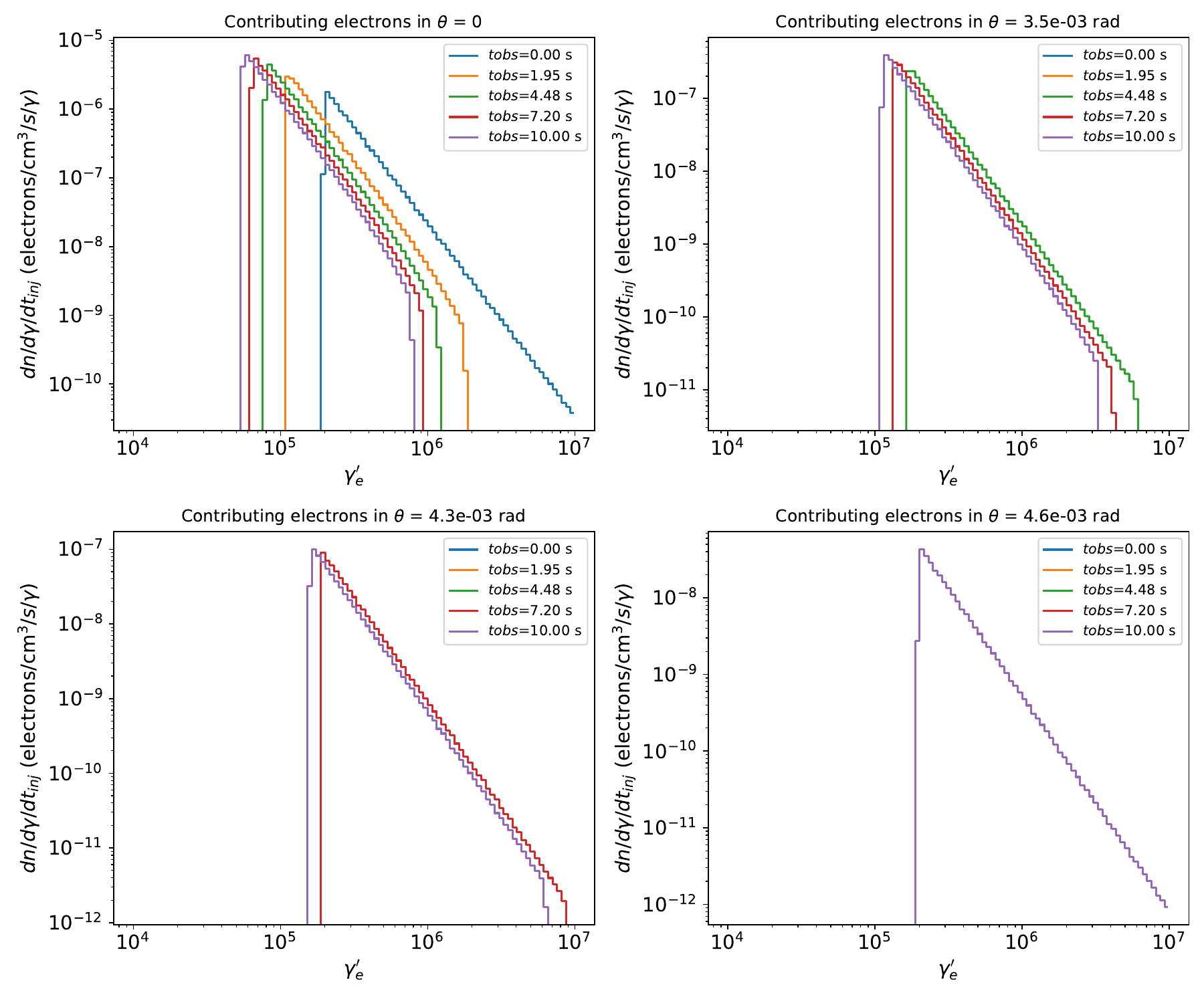}
    \caption{The evolution of electron spectrum at different observer's time since injection from different latitudes. The parameters are listed in section \ref{sec:reproduce}.}
    \label{fig:dndg}
\end{figure}

After $dn_e(\gamma^\prime_e, t; t_\theta)/d\gamma^\prime_e$ is known, we integrate the synchrotron radiation contributed by electrons at each $\gammae$:
\begin{equation}
    f(\nuem^\prime,t;t_\theta)= \Delta r^\prime\int^{\gamma^\prime_M}_{\gamma^\prime_m} P_{\text{syn}}(\nuem^\prime, B^\prime, \gammae)\frac{dn_e(\gamma^\prime_e, t; t_\theta)}{d\gamma^\prime_e}d\gammae, \label{eq:emissivity}
\end{equation}
where the synchrotron radiation power for single electron at $\gammae$ is:
\begin{equation}
    P_{\text{syn}}(\nuem^\prime, B^\prime, \gammae)=  \frac{\sqrt{3} e^3 B^\prime}{m_{\mathrm{e}} c^2} \cdot \frac{\nuem^\prime}{\nu_c} \cdot \int_{\nu/\nu_c}^{\infty} \mathrm{K}_{5/3}(x) \, \mathrm{d}x
\end{equation}
where $\text{K}_{5/3}$ denotes the modified Bessel function, and $\nu_c$ is the critical frequency related with $\gammae$:
\begin{equation}
    \nu_c = \frac{3 e B^{\prime}{\gammae}^2}{4 \pi m_{\mathrm{e}} c}.
\end{equation}
$\Delta r^\prime$ is the thickness of the shell of dissipation, in our setting, $\Delta r^\prime\ll R/\Gamma$, corresponding to a thin shell in the comoving frame. 

Taking equation (\ref{eq:emissivity}) into equation (\ref{eq:fv}) allows us to obtain the observed specific flux as a function of time and photon energy. 

\section{A {\revise new} catalogue of general single FRED GRBs from {\it Fermi}/GBM}
{\revise In BASTE's era, catalogues of single FRED GRBs have been presented, under the name of ``long-lag wide-pulse GRBs'', most notably \citep{2005ApJ...627..324N}}. As an extension, here we present a {\revise new} catalogue of single FRED GRBs found in {\it Fermi}/GBM GRBs {\revise as an extension to this population}, and find their energy-dependence of pulse width and peak time. 

To begin with, we extracted the lightcurves in four energy channels (i.e. 8-30\,keV, 30-100\,keV, 100-300\,keV, 300-900\,keV) for all cataloged \textit{Fermi}/GBM GRBs \citep{GBM_catalog} using the archived Time-Tagged Event (TTE) data of all NaI detectors. Then the searching for single FRED sample is conducted by the following step:
\begin{enumerate}[(1)]
    \item We restrict our sample to GRBs with $T_{90}>2\,$s, as they are expected to exhibit more pronounced spectral lags, enabling a more robust determination of the energy dependence law. GRBs with a fluence less than $10^{-5}$\,erg$\cdot$cm$^{-2}$ were also excluded, to ensure sufficient counts for reliable analysis.
    \item The lightcurve of 300-900\,keV (the hardest energy range we used) is rebinned with a time resolution of 0.5\,s from T$_0$-T$_{90}$ to T$_0$+3T$_{90}$, where T$_0$ set as the trigger time for each GRBs. 
    \item Then this rebinned 300-900\,keV lightcurve is fitted by a FRED model \citep{2005ApJ...627..324N}, which is stated as Equation \ref{equ:norris05}, superposed on a one-order polynomial background by the same MCMC approach with \cite{2025ApJ...985..239Y}. GRBs with residuals less than 5 are selected as part of single FRED GRBs sample. 
    \item The rebinned 300-900\,keV lightcurve in step (2) is also used to perform peak searching by MEPSA \citep{MEPSA}. GRBs with only one pulse identified are contributed to the other part of the single FRED GRBs sample.
\end{enumerate}
A total of 28 single pulse GRBs are collected based on the above steps. 
Then we fitted the four channel lightcurves of these 28 GRBs with the same process in step (3). We set a prior of the parameter imposing that both the rise time scale and decay time scale should shorter than T$_{90}$. We excluded GRBs whose parameters are not well constrained, and left 17 GRBs in our final sample, {\reviseJHEAp which are shown in Figure\,\ref{fig:lc_catalog_1} to Figure\,\ref{fig:lc_catalog_3}}.

The pulse width and peak time are deduced based on the parameters of FRED with the same definition in \cite{2025ApJ...985..239Y}, and fitted by powerlaw to get their energy-dependence (i.e. the slope) respectively, as shown in the figure \ref{fig:pop-tw}. {\reviseJHEApB As we can see in GRB 130304A and GRB 161206A (panels a and e of Figure\,\ref{fig:lc_catalog_1}), there are fast pulses resolved from the overall FRED-like profiles. Such features were also clearly manifested in GRB 230307A, where the overall FRED profile was shown to be composed of multiple fast pulses. In our framework, these features correspond to stochastic local reconnection events sequentially triggered by the propagating perturbation within a single epoch of central engine activity. }

\begin{table*}[htbp]
\caption{\centering{Sample of general single FRED GRBs}}
\begin{tabular*}{\hsize}{@{}@{\extracolsep{\fill}}cccc@{}}
\toprule
GRB name & T$_0$ & Slope of t$_p$ & Slope of t$_w$ \\
\hline
 GRB 130304A & 2013-03-04T09:49:53.099 & -0.37$^{+0.18}_{-0.16}$ & -0.33$^{+0.03}_{-0.03}$\\ 
 GRB 130704A & 2013-07-04T13:26:07.253 & -0.42$^{+0.00}_{-0.00}$ & -0.26$^{+0.00}_{-0.00}$\\ 
 GRB 140723A & 2014-07-23T01:36:30.728 & -0.33$^{+0.07}_{-0.07}$ & -0.55$^{+0.04}_{-0.04}$\\ 
 GRB 160101A & 2016-01-01T00:43:53.610 & -0.12$^{+0.01}_{-0.01}$ & -0.12$^{+0.01}_{-0.01}$\\ 
 GRB 161206A & 2016-12-06T01:32:28.077 & -0.14$^{+0.01}_{-0.01}$ & -0.12$^{+0.01}_{-0.01}$\\ 
 GRB 170921B & 2017-09-21T04:02:11.511 & -0.93$^{+0.01}_{-0.01}$ & -0.38$^{+0.01}_{-0.01}$\\ 
 GRB 171210A & 2017-12-10T11:49:15.261 & -0.26$^{+0.01}_{-0.01}$ & -0.93$^{+0.01}_{-0.01}$\\ 
 GRB 180426A & 2018-04-26T13:11:00.846 & -0.03$^{+0.03}_{-0.03}$ & -0.25$^{+0.03}_{-0.03}$\\ 
 GRB 180806A & 2018-08-06T22:38:59.661 & -0.13$^{+0.01}_{-0.01}$ & -0.08$^{+0.01}_{-0.01}$\\ 
 GRB 190222A & 2019-02-22T12:53:27.151 & -0.46$^{+0.04}_{-0.04}$ & -0.45$^{+0.03}_{-0.03}$\\ 
 GRB 190604A & 2019-06-04T10:42:37.054 & -0.14$^{+0.01}_{-0.01}$ & -0.14$^{+0.01}_{-0.01}$\\ 
 GRB 200607B & 2020-06-07T22:06:31.425 & -0.17$^{+0.14}_{-0.14}$ & -0.77$^{+0.03}_{-0.02}$\\ 
 GRB 211116A & 2021-11-16T14:03:53.348 & -0.36$^{+0.03}_{-0.03}$ & -0.39$^{+0.02}_{-0.02}$\\ 
 GRB 221221A & 2022-12-21T22:39:30.568 & -0.19$^{+0.03}_{-0.03}$ & -0.19$^{+0.02}_{-0.02}$\\ 
 GRB 230402A & 2023-04-02T07:32:35.956 & -0.24$^{+0.02}_{-0.02}$ & -0.59$^{+0.02}_{-0.02}$\\ 
 GRB 230621A & 2023-06-21T23:45:24.826 & -0.17$^{+0.04}_{-0.04}$ & -0.35$^{+0.03}_{-0.03}$\\ 
 GRB 240914B & 2024-09-14T07:09:47.313 & -0.10$^{+0.01}_{-0.01}$ & -0.10$^{+0.01}_{-0.01}$\\ 
\botrule
\end{tabular*}
\end{table*}

\clearpage
\begin{figure*}
\centering
\begin{tabular}{cc}
\begin{overpic}[width=0.5\textwidth]{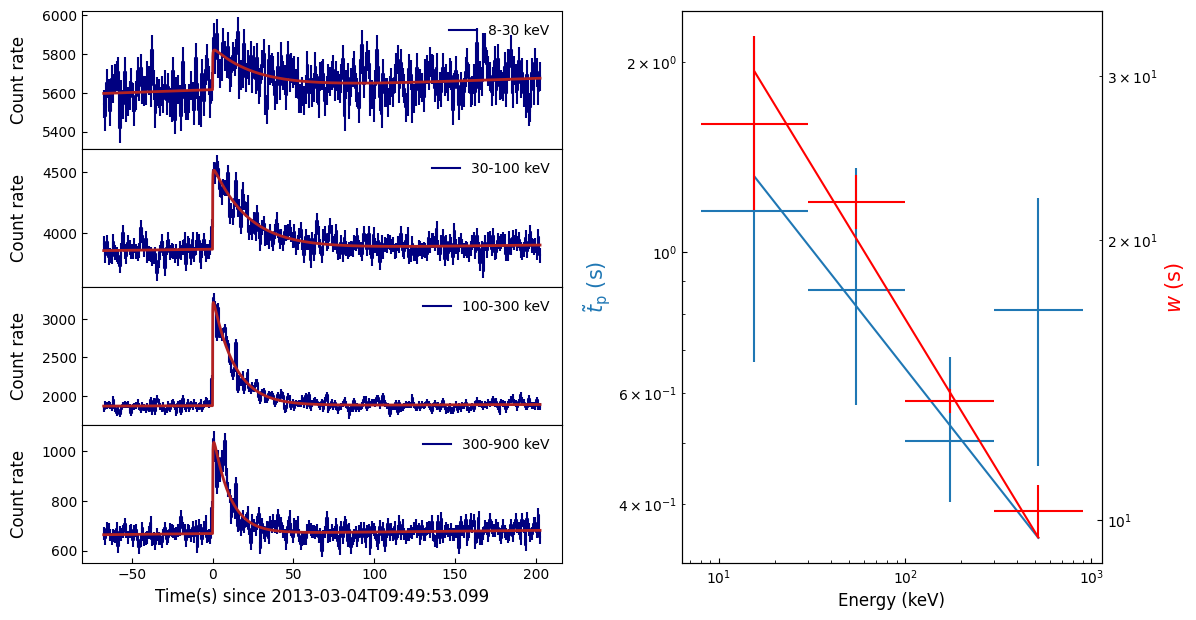}\put(-3, 50){\bf a}\put(45,1){\bf \tiny GRB 130304A}\end{overpic}&
\begin{overpic}[width=0.5\textwidth]{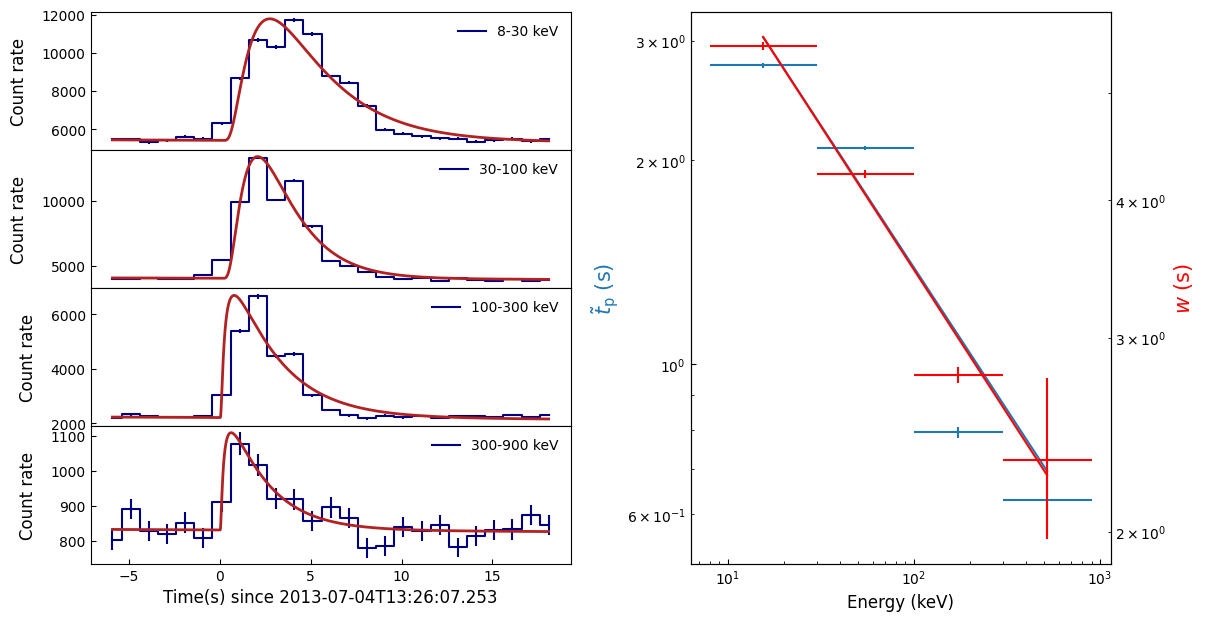}\put(-3, 50){\bf b}\put(45,1){\bf \tiny GRB 130704A}\end{overpic}\\
\begin{overpic}[width=0.5\textwidth]{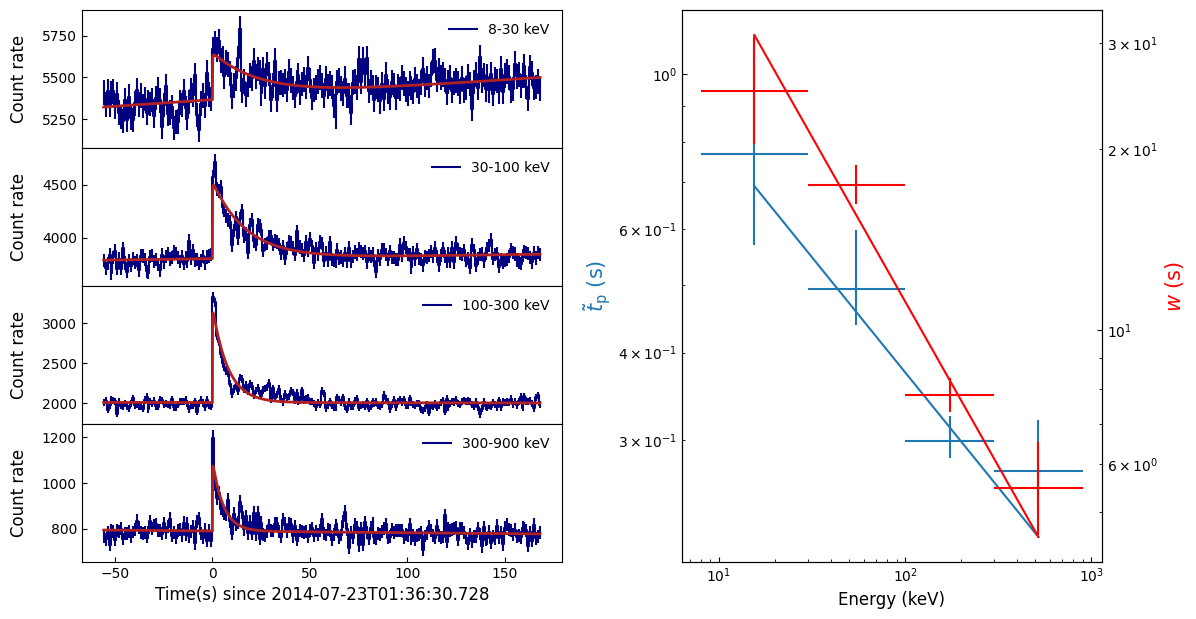}\put(-3, 50){\bf c}\put(45,1){\bf \tiny GRB 140723A}\end{overpic}&
\begin{overpic}[width=0.5\textwidth]{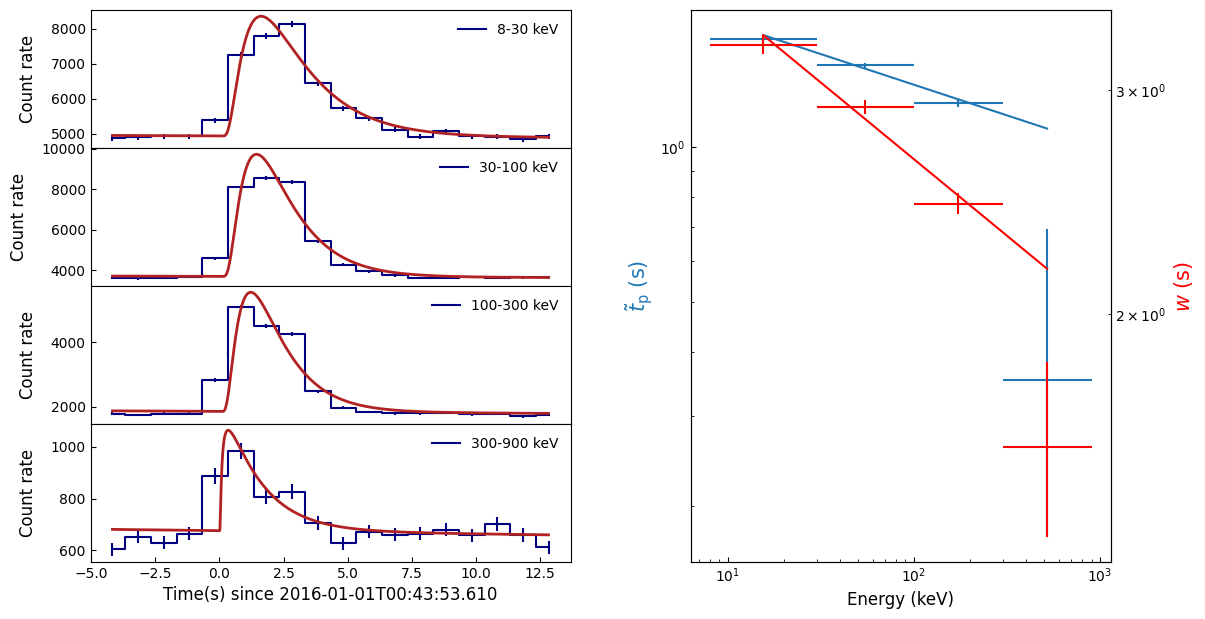}\put(-3, 50){\bf d}\put(45,1){\bf \tiny GRB 160101A}\end{overpic}\\
\begin{overpic}[width=0.5\textwidth]{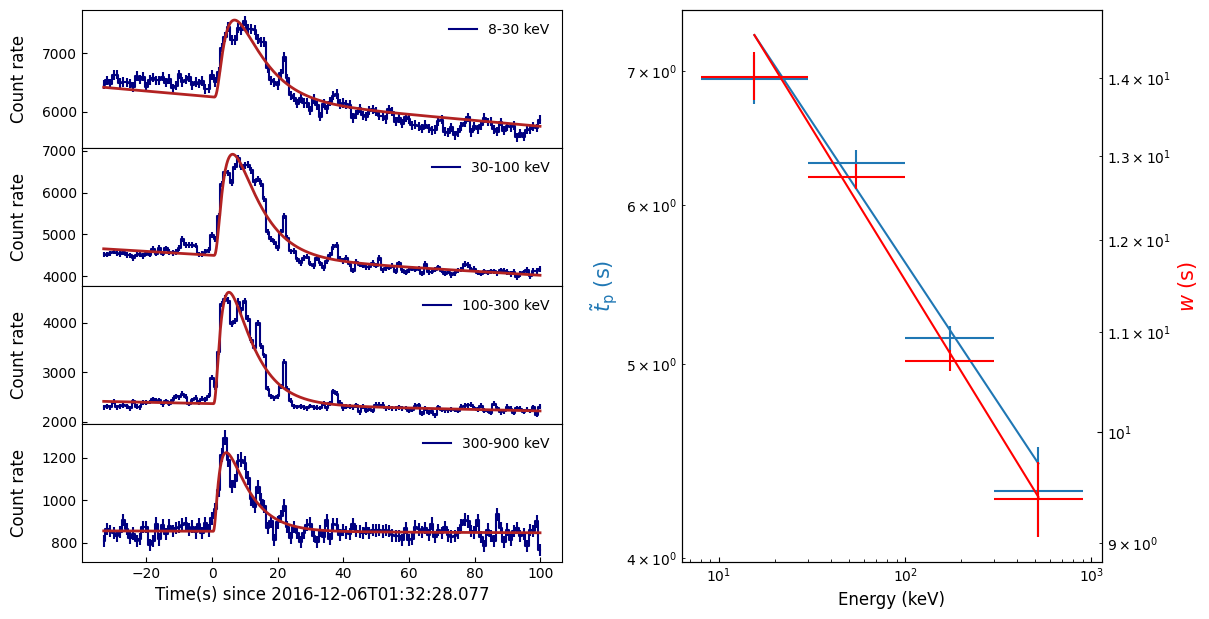}\put(-3, 50){\bf e}\put(45,1){\bf \tiny GRB 161206A}\end{overpic}&
\begin{overpic}[width=0.5\textwidth]{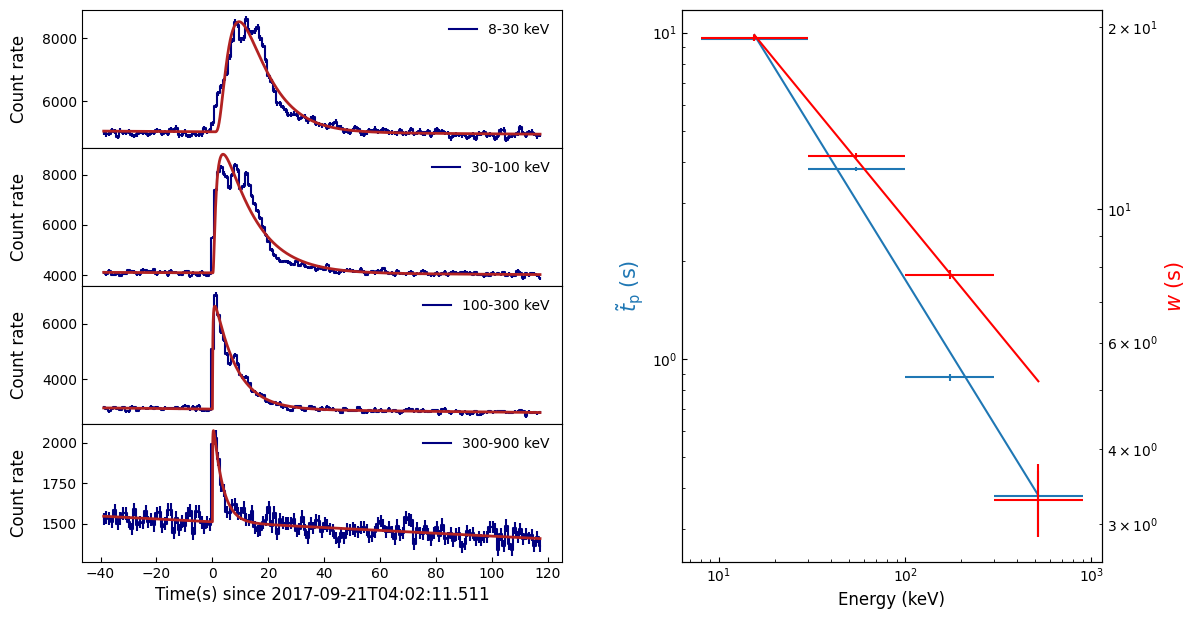}\put(-3, 50){\bf f}\put(45,1){\bf \tiny GRB 170921B}\end{overpic}\\
\begin{overpic}[width=0.5\textwidth]{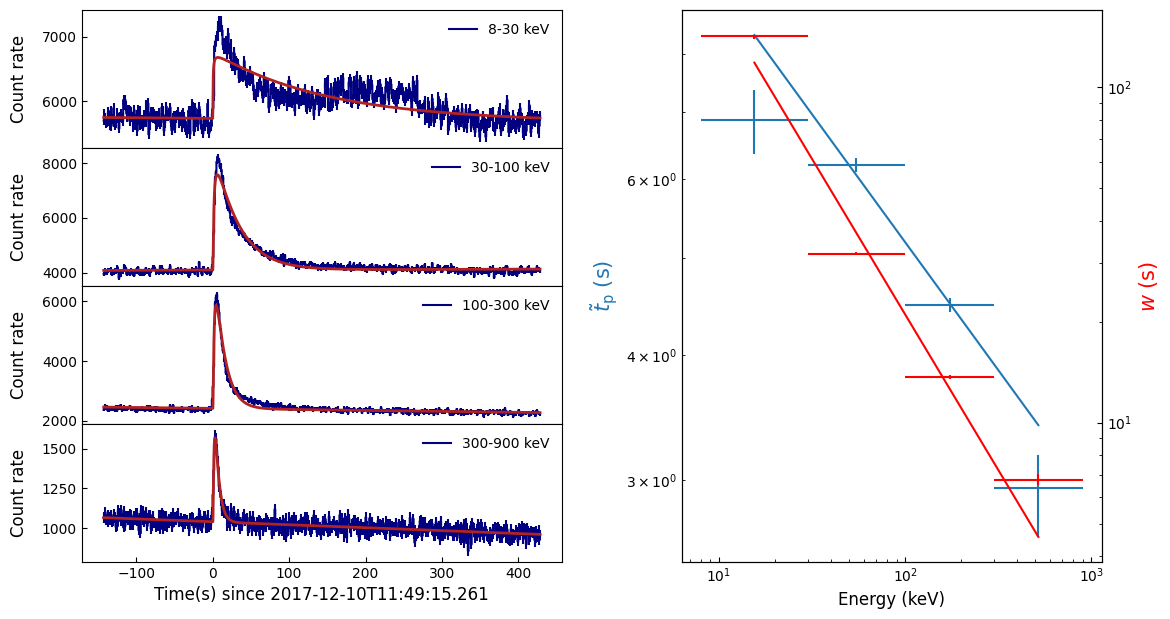}\put(-3, 50){\bf g}\put(45,1){\bf \tiny  GRB 171210A}\end{overpic}&
\begin{overpic}[width=0.5\textwidth]{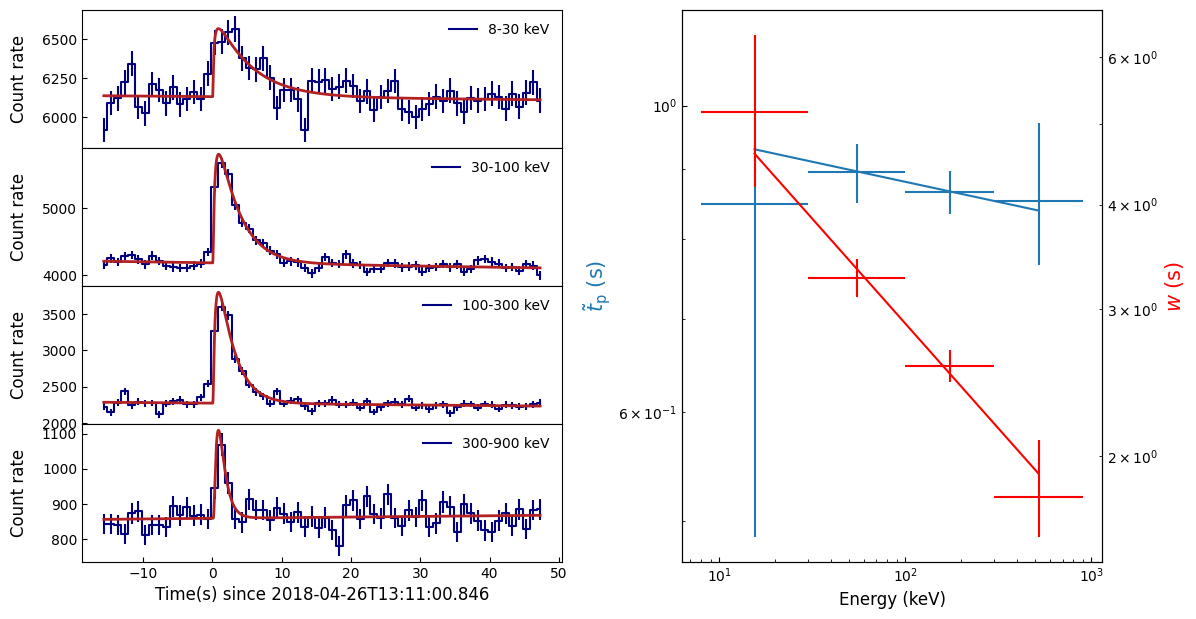}\put(-3, 50){\bf h}\put(45,1){\bf \tiny GRB 180426A}\end{overpic}
\end{tabular}
\caption{
The light curves of the new general single FRED GRBs catalog. The names of the GRBs are indicated in the bottom of each panel.
}
\label{fig:lc_catalog_1}
\end{figure*}

\begin{figure*}
\centering
\begin{tabular}{cc}
\begin{overpic}[width=0.5\textwidth]{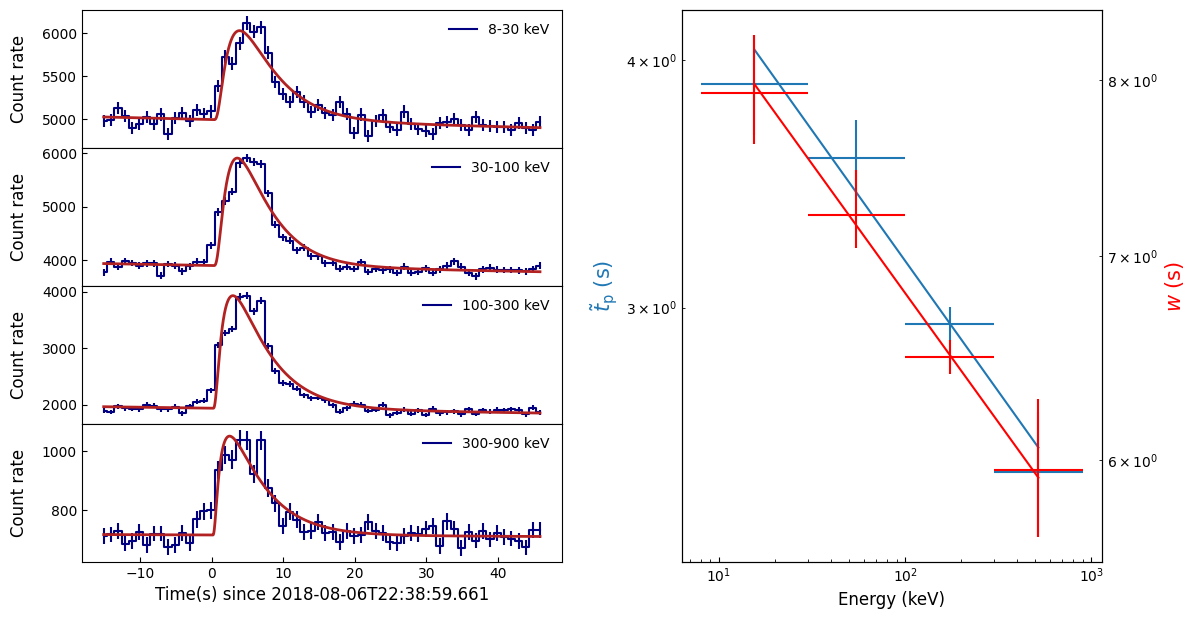}\put(-3, 50){\bf i}\put(45,1){\bf \tiny GRB 180806A}\end{overpic}&
\begin{overpic}[width=0.5\textwidth]{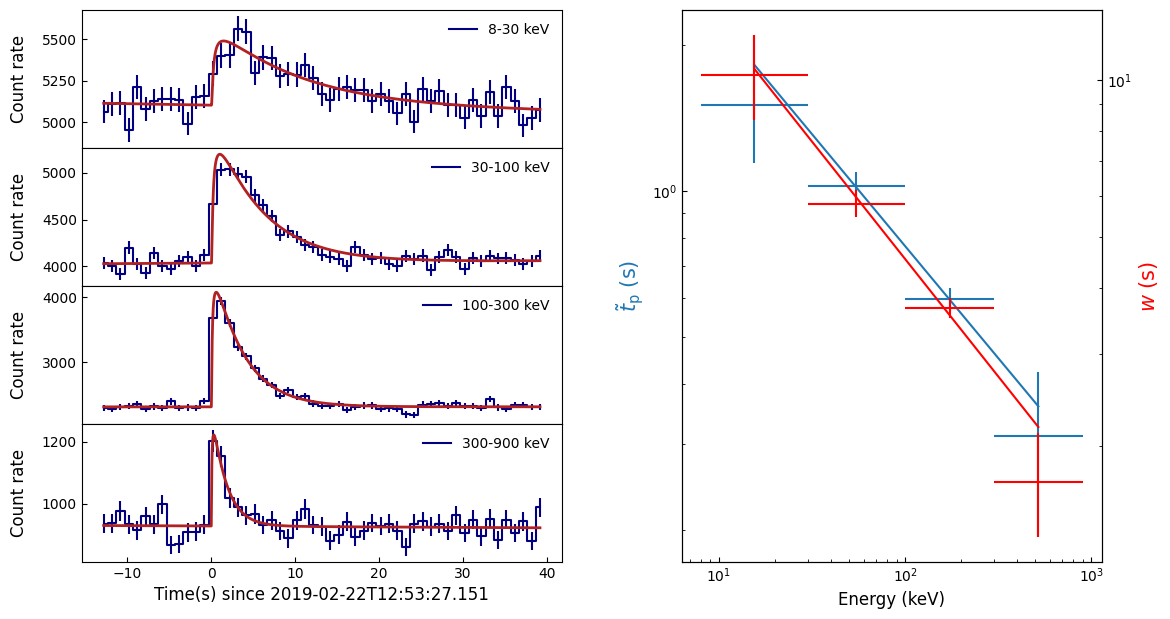}\put(-3, 50){\bf j}\put(45,1){\bf \tiny GRB 190222A}\end{overpic}\\
\begin{overpic}[width=0.5\textwidth]{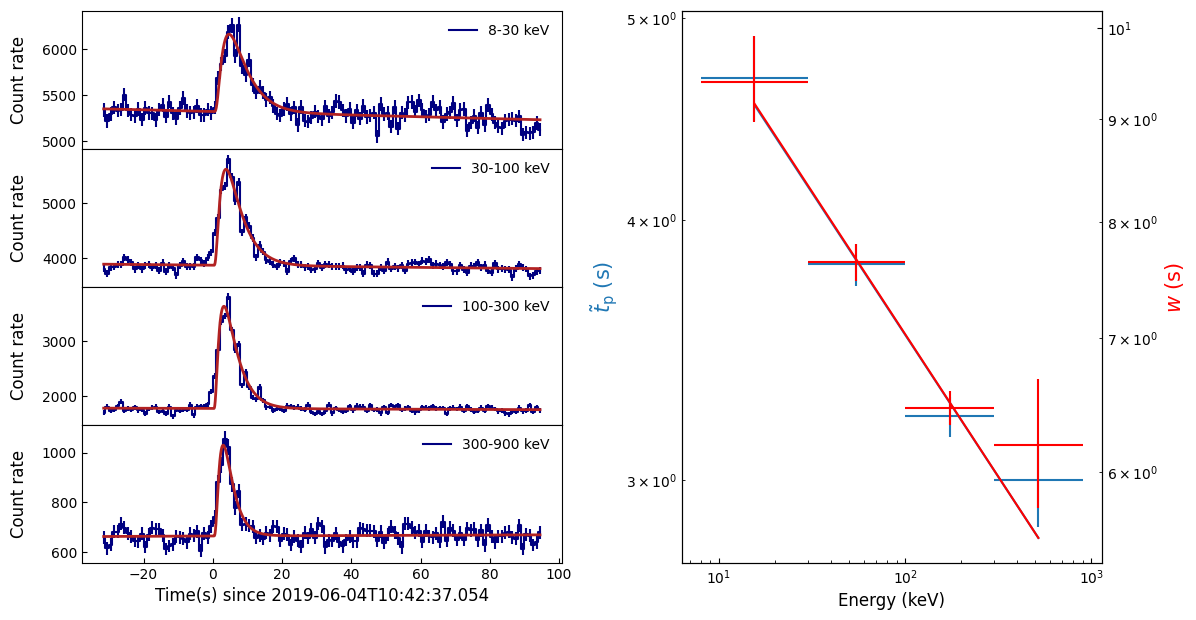}\put(-3, 50){\bf k}\put(45,1){\bf \tiny GRB 190604A}\end{overpic}&
\begin{overpic}[width=0.5\textwidth]{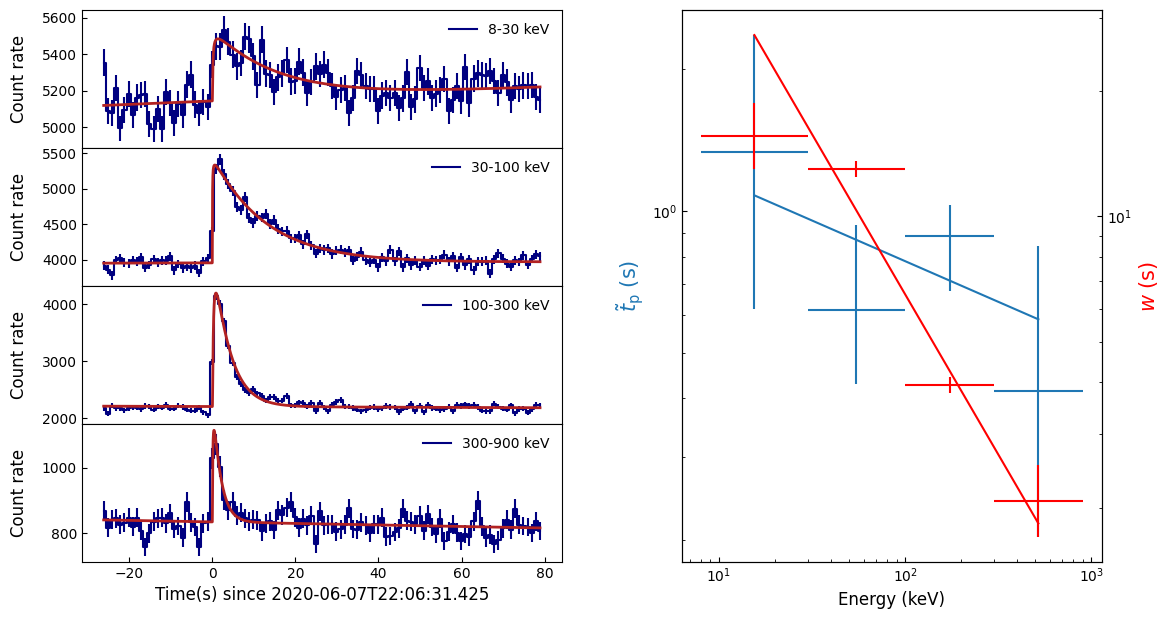}\put(-3, 50){\bf l}\put(45,1){\bf \tiny GRB 200607B}\end{overpic}\\
\begin{overpic}[width=0.5\textwidth]{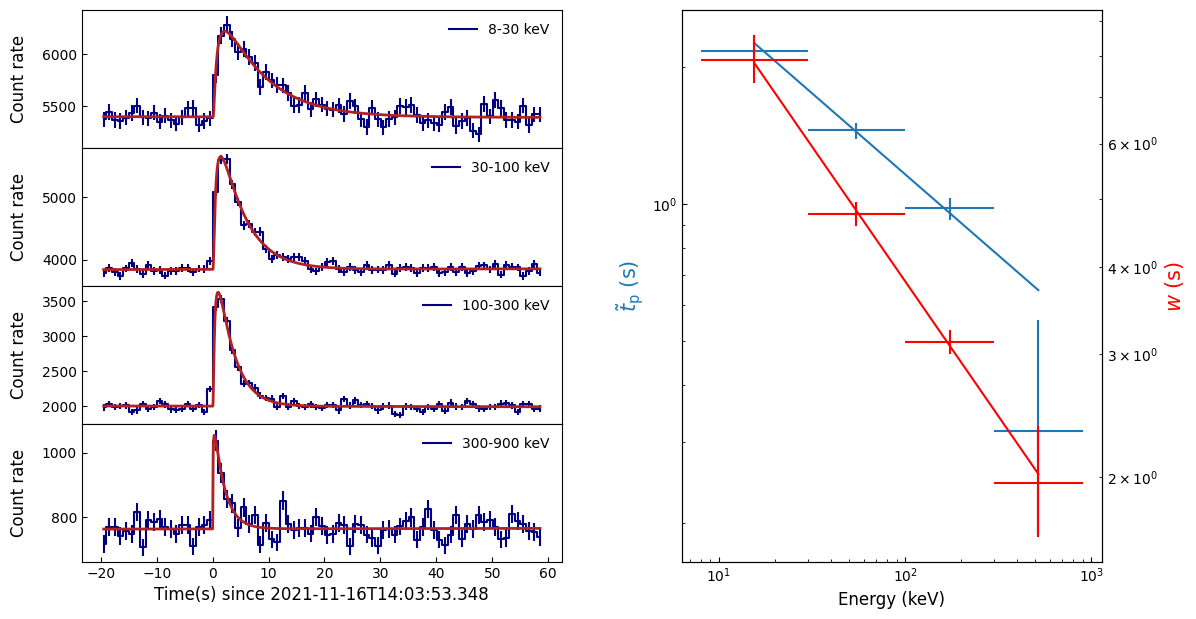}\put(-3, 50){\bf m}\put(45,1){\bf \tiny GRB 211116A}\end{overpic}&
\begin{overpic}[width=0.5\textwidth]{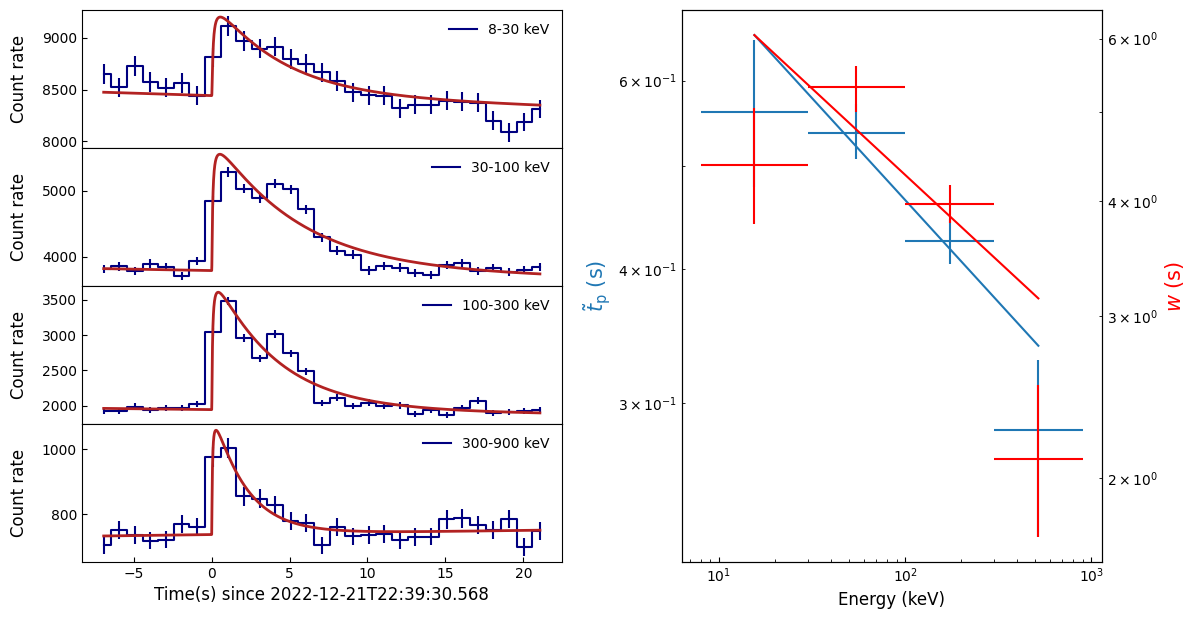}\put(-3, 50){\bf n}\put(45,1){\bf \tiny GRB 221221A}\end{overpic}\\
\begin{overpic}[width=0.5\textwidth]{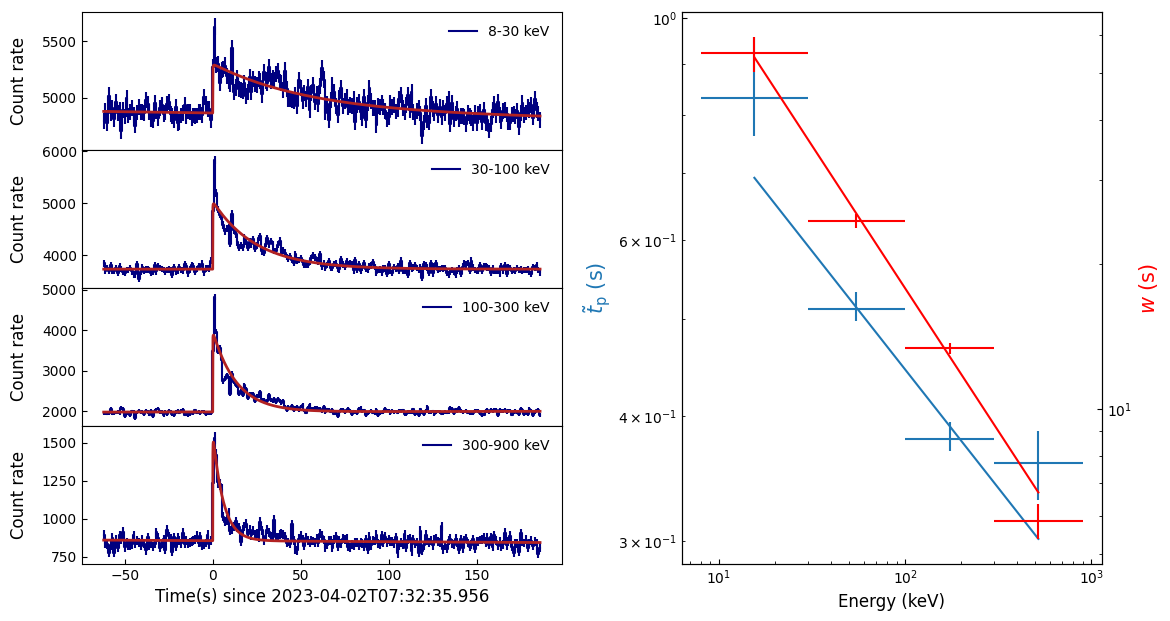}\put(-3, 50){\bf o}\put(45,1){\bf \tiny GRB 230402A}\end{overpic}&
\begin{overpic}[width=0.5\textwidth]{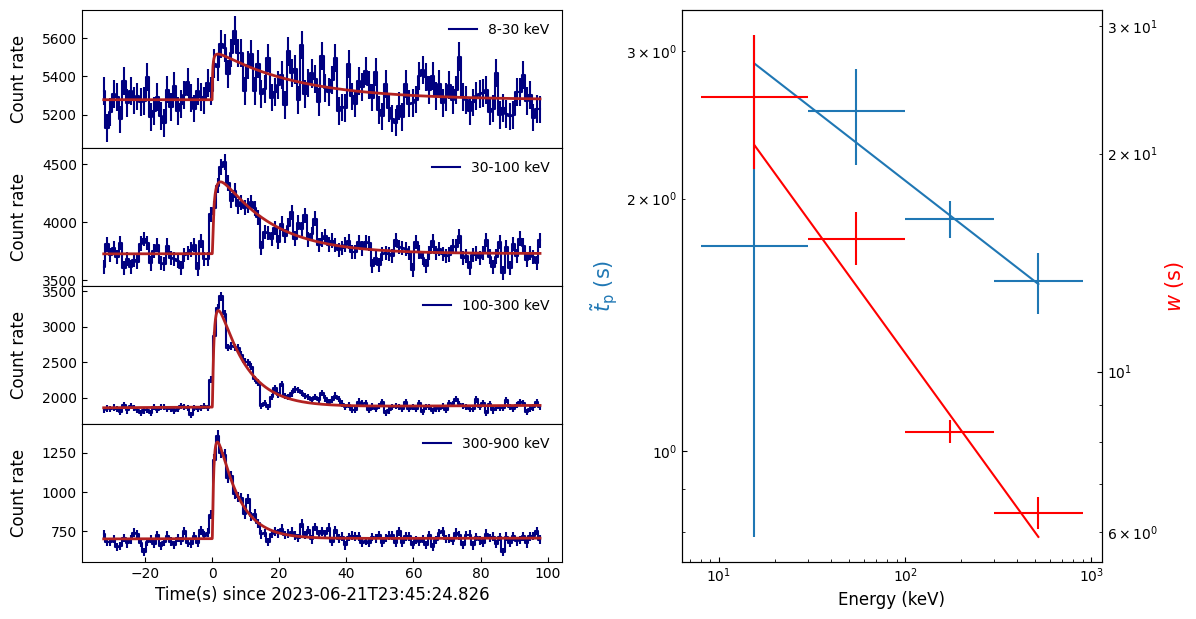}\put(-3, 50){\bf p}\put(45,1){\bf \tiny GRB 230621A}\end{overpic}
\end{tabular}
\caption{
(Continue) The light curves of the new general single FRED GRBs catalog. 
}
\label{fig:lc_catalog_2}
\end{figure*}

\begin{figure*}
\centering
\begin{tabular}{cc}
\begin{overpic}[width=0.5\textwidth]{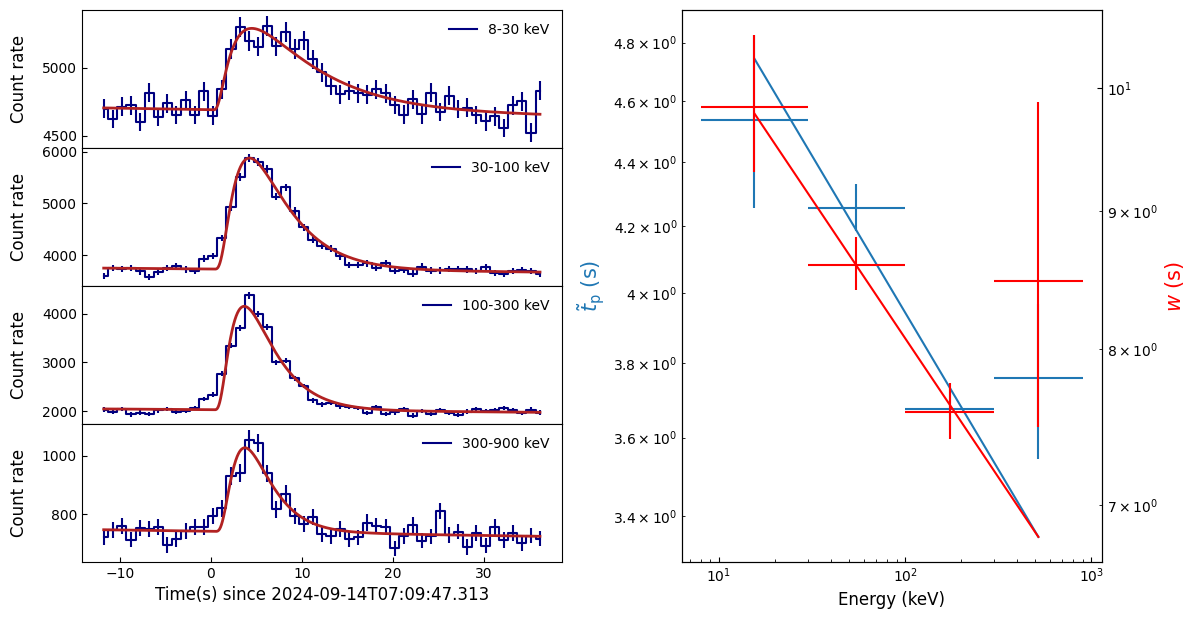}\put(-3, 50){\bf q}\put(45,1){\bf \tiny GRB 240914B}\end{overpic}
\end{tabular}
\caption{
(Continue) The light curves of the new general single FRED GRBs catalog. 
}
\label{fig:lc_catalog_3}
\end{figure*}

\clearpage
\bibliography{ref}{}
\end{document}